\documentclass[numbers,3p, times]{elsarticle}
\usepackage{lineno,hyperref}
\modulolinenumbers[5]
\usepackage{subcaption}

\journal{Journal of }
\usepackage{supertabular}
\usepackage{amsmath}
\usepackage{amsbsy}
\usepackage{amssymb}
\usepackage{multicol}

\usepackage{algorithm}
\usepackage{algorithmic}
\usepackage{multirow}
\usepackage{xcolor}

\usepackage[english]{babel}

\usepackage[figuresright]{rotating}

\usepackage[normalem]{ulem}

\newcommand{\field}[1]{\mathbb{#1}}
\newcommand{\R}{\field{R}}

\begin{document}

\begin{frontmatter}

\title{A surrogate model for studying random field energy release rates in 2D brittle fractures
} 

\author[label1]{Luis Blanco-Cocom\corref{cor1}}
\author[label2]{Marcos A. Capistr\'an}
\author[label3]{Jaroslaw Knap}
\author[label1]{J. Andr\'es Christen}

 \address[label1]{Centro de Investigaci\'on en Matem\'aticas, A.C., Unidad Guanajuato, S/N, Col. Valenciana CP: 36240 Guanajuato, Gto, M\'exico, Apartado Postal 402, CP 36000 }
  \address[label2]{Centro de Investigaci\'on en Matem\'aticas A.C. Unidad M\'erida, km 5.5 Carretera Sierra Papacal - Chuburn\'a Puerto, M\'erida 97302, Yucat\'an, M\'exico}
  \address[label3]{U.S. Army Research Laboratory, Aberdeen Proving Ground, MD 21005, USA.}

 \cortext[cor1]{Corresponding author: luis.blanco@cimat.mx}

\begin{abstract}
This article proposes a weighted-variational model as an approximated surrogate model to lessen numerical complexity and lower the execution times of brittle fracture simulations. Consequently, Monte Carlo studies of brittle fractures become possible when energy release rates are modelled as a random field. In the weighed-variational model, we propose applying a Gaussian random field with a Mat\'ern covariance function to simulate a non-homogeneous energy release rate ($G_c$) of a material. Numerical solutions to the weighed-variational model, along with the more standard but computationally demanding hybrid phase-field models, are obtained using the FEniCS open-source software. The results have indicated that the weighted-variational model is a competitive surrogate model of the hybrid phase-field method to mimic brittle fractures in real structures. This method reduces execution times by 90\%.  We conducted a similar study and compared our results with an actual brittle fracture laboratory experiment. We present an example where a Monte Carlo study is carried out, modeling $G_c$ as a Gaussian Process, obtaining a distribution of possible fractures, and load-displacement curves.
\end{abstract}


\end{frontmatter}

\section{Introduction}

Over the years, there has been substantial focus on brittle fracture simulation in multiple scientific fields. The primary objective of brittle fracture modeling is to forecast the timing and the properties of a brittle fracture, inclusive of its trajectory, velocity, branching, and coalescence~\cite{MIEHE_2010,BLEYER_2018,TANNE_2018,ULLOA_2019}.
The classical Griffith theory of brittle fracture provides a criterion for crack propagation, yet it is alone incapable of determining crack trajectories or crack branching~\cite{MIEHE_2010,CHARLOTTE_2006}. This has motivated the development of numerous new methods and methodologies. In the literature, several mathematical models attempt to describe the propagation of cracks, an isotropic second-order phase-field fracture model was introduced in \cite{FRANCFORT_1998}, also an anisotropic model derived from the model in \cite{BOURDIN_2000} is studied in \cite{AMOR_2009}; the latter considers the volumetric and deviatoric components in the elastic energy density to prevent cracking in the compression domain. A fourth-order phase field model to improve convergence rates for numerical solutions using isogeometric finite elements is analyzed in \cite{BORDEN_2014}. In \cite{Yosra_2020} a matrix assembly optimization method for the FEM algorithm with refined mesh is introduced to decrease computational expenses in simulations of the hybrid phase-field model.  

In the last decade, many phase-field brittle fracture models have been introduced, yielding remarkably accurate predictions to simulate brittle fractures for various material classes and loading scenarios~\cite{MIEHE_2010,BORDEN_2012,HESCH_2014,WEINBERG_2017}; see \cite{WU_2020} for a review on phase-field modeling.

The popularity of the phase-field method stems from the fact that conventional numerical methods, such as finite elements, can be easily employed to seek solutions of phase-field equations, arising from the fact that the cracks are represented only by a continuous field, as explained in~\cite{HIRSHIKESH_2019}. 

Recently, methods based on variational approaches have been studied~\cite{FRANCFORT_1998,BOURDIN_2008,MASO_2002,BULIGA_1998}.  Within this variational framework, the process of brittle fracture is fully characterized by directly minimizing an energy functional~\cite{BOURDIN_2008, MESGARNEJAD_2015,AREIAS_2016}, and the numerical methods may be systematically derived through $\Gamma$-convergence analysis~\cite{MUNFORD_1989}.

Models of brittle fracture often assume spatial homogeneity of the underlying material properties. This assumption is, however, seldom valid as the microstructure underlying material behavior tends to be endowed with randomness due to the presence of defects, such as grains, inclusions, voids, etc. Hence, for modeling brittle fractures, non-homogeneous random properties must be included.
A few brittle fracture models directly incorporating randomness have been proposed.  For instance  in \cite{YANG_2008}, a heterogeneous cohesive (HC) crack model is developed to predict macroscopic strength of materials based on meso-scale random fields of fracture properties. Moreover, a study of potential cracks represented by pre-inserted cohesive elements with tension and shear, softening constitutive laws, are modeled by spatially-varying Weibull random fields in \cite{YANG_2009,SU_2010}. A statistical analysis of graded materials was performed in a recent study by \cite{Dsouza_2021}, using a non-intrusive polynomial chaos expansion to investigate the randomness of Young's modulus, fracture toughness, and gradient index. 

%\cutmc{In general, mathematical models present a computational complexity that depends on the characteristics to be studied or of interest. Simpler mathematical models have been used to analyze behaviors in simulations without loss of precision or accuracy; these models are called surrogate models \cite{Asher_2015}.  This models allow carrying out simulations in less time, can be calibrated, and present fewer numerical instabilities compared to complex models. These kind of models have been applied in several branches of the science \cite{CAI_2021,Lan_2020,Chris_2019,Asher_2015,Tirnovan_2008}.}

In most mechanical systems, specifically brittle fracture models, there exists a direct correlation between model accuracy and computational expense. Frequently, it is only computationally feasible to perform a small number of simulations with high precision. As a result, numerous research endeavors are focused on creating Fidelity-Reduced Order surrogate Models (ROMs)~\cite{CAI_2021,Lan_2020,Chris_2019,Asher_2015,Tirnovan_2008} or multiple-fidelity models~\cite{fernandez2023review,peherstorfer2018survey,ALLAIRE_2014}, or approaches using Neural Networks \cite{GOSWAMI_2020,Aydin_2019}. Although these approaches have been successful, issues regarding the development and usage of both multi-fidelity and surrogate models remain unsolved~\cite{giselle2019issues}.  

To accurately reproduce a solution to a real-world problem, a mathematical model involves a complex formulation to maintain a high fidelity to the properties of reality and requires multiple formulas or long runtimes for simulations. Multifidelity models combine multiple fidelities in a single model to achieve the desired level of accuracy at a lower cost \cite{fernandez2023review}, these models require the construction of surrogate models to reduce the computational complexity associated with a large number of expensive simulations. This is particularly useful for tasks such as optimization and uncertainty quantification.

This article proposes the use of a Gaussian random field with a Mat\'ern covariance function to model the spatial non-homogeneity of the energy release rate $G_c$. Inclusion of non-homogeneity transforms the scalar $G_c$ into a random field, significantly increasing model complexity.  A Monte Carlo study would be unfeasible without a fast and reliable surrogate model.
In turn, we propose using a computationally cheaper surrogate model that allows multiple simulations of crack propagation under Monte Carlo samples of the Gaussian random field that governs $G_c$. This method will permit the examination of the resulting probability distribution of fracture trajectories and load-displacement curves.  

Our proposal is to utilize a weighted-variational technique as an affordable surrogate model to enable Monte Carlo analyses of fracture paths, 
particularly when we assume that the energy release rate $G_C$ is distributed as a Gaussian random field. The surrogate model proposed in this paper is based on partially converged solutions of the well-established variational phase-field fracture model presented in \cite{BOURDIN_2000}, and
provides 
%Our proposal of the weighted-variational model, 
reasonable surrogate simulations to substitute the hybrid phase-field model, reducing execution times by 90\% for the selected examples.  The weighted-variational and hybrid phase-field model equations are solved by implementing a code in Python with the open-source finite element software FEniCS as in ~\cite{TANNE_2018} and ~\cite{HIRSHIKESH_2019}.

The paper is organized as follows: Section \ref{sec:2} presents an overview of the hybrid phase-field and weighted-variational mathematical models. Details of the FEniCS implementation of both
models are presented in Section \ref{sec:4}.  In section \ref{sec:3}, the formulation of the Gaussian random field to model the energy release rate using a Mat\'ern covariance matrix is presented. In Section \ref{sec:5}, we present the solution of the standard boundary value problems to validate the implementation in FEniCS.  We also show simulations using the Gaussian random fields for $G_c$, and present comparisons of our proposed surrogate model with a laboratory experiment.  A comprehensive Monte Carlo study is presented, simulating 500 crack trajectories, to estimate quantiles of the crack probability distribution and load-displacement curves. Finally, a discussion of the paper is presented in Section \ref{sec:6}.

\section{Mathematical models}\label{sec:2}

In this section, we introduce the hybrid phase field mathematical model for brittle fracture propagation and the
weighted-variational model proposed as a computationally faster surrogate for the former.  See Table~\ref{table_vars} for a list of the main parameters used.

\begin{table*}[ht!]
\indexspace
%\begin{minipage}{\linewidth}
\begin{center}
\caption{List of main parameters used.}
\label{table_vars}
\begin{tabular}{|ll|}
\hline 
$\boldsymbol{u}$ & Displacement field\\
$\phi$ & Scalar damage variable\\
$\lambda$, $\mu$ & Lamm\'e constants (kN/mm$^2$)\\
$G_c$ & Critical energy release rate (kN/mm)\\
$\eta_{\ell}$ & Constant of order $O(\ell)$ \\
$\ell$ & Regularization length (mm)\\
$\psi$ & Elastic energy density\\
$\boldsymbol{\varepsilon}$ & Small strain tensor\\
$H$ & History variables \\
$H_n$ & Strain energy computed at load step n\\
$\Delta u$ & Incremental displacement (mm)\\
$\overline{\boldsymbol{t}}$ & Traction (kN/mm$^2$) \\
$C(\cdot)$ & Multidimensional covariance function \\
$S(\cdot)$ & Spectral density function \\
$\mathcal{F}$ & Fourier transform \\
\hline
\end{tabular}
\end{center}
\end{table*}

\subsection{The hybrid phase field method}
\begin{figure}[H]
    \begin{center} 
  \includegraphics[width=8 cm]{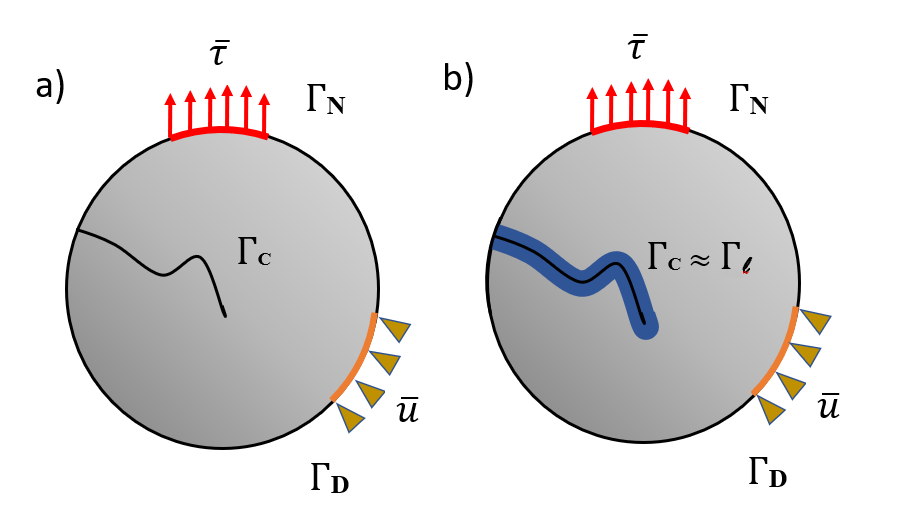}
\end{center} 
\caption{Specimen $\Omega$ with an inner crack $\Gamma_C$: a) sharp crack, b) approximated diffuse crack as function of $\ell$.}\label{body_1}
\end{figure}

Recently, the hybrid phase-field mathematical model for brittle fracture has been used to simulate brittle cracks \cite{HIRSHIKESH_2019,AMBATI_2015}, this mathematical approach has its foundations in the modification of the variational model presented in \cite{MIEHE_2010}; see  \cite{GULTEKIN_2016} for details.
The resulting system consists of two coupled non-linear differential equations, the
first equation describes the phase-field to emulate the fracture, and the second equation describes the displacements on the domain $\Omega$. The phase field system is described by
\cite{HIRSHIKESH_2019},
{
\begin{equation}\label{sis_1}
\left\lbrace\begin{array}{ll} 
 \nabla \cdot \boldsymbol{\sigma} = 0, & \text{in } \quad \Omega, \\
 -G_c \ell \nabla ^2 \phi + \left[ \frac{G_c}{\ell} + 2 H \right] \phi = 2H, & \text{in } \quad \Omega, \\
% G_c\left[ -\ell^2 \nabla^2 \phi + \phi \right] = 2\ell (1-\phi)H , & \text{in } \quad \Omega,
\end{array}\right. \\
\end{equation}
}
 with boundary conditions,

\begin{equation}
\left\lbrace\begin{array}{ll} 
 \boldsymbol{n}  \cdot \boldsymbol{\sigma} = \overline{\boldsymbol{t}}, & \text{on } \quad \Gamma_N, \\
 \boldsymbol{u} = \overline{\boldsymbol{u}}, & \text{on } \quad \Gamma_D, \\
 \nabla \phi \cdot \boldsymbol{n} = 0, & \text{on } \quad \Gamma_N,
\end{array}\right.
\end{equation}
where {$\boldsymbol{\sigma} = \left[ (1 - \phi)^2 + \eta_{\ell} \right] \frac{\partial \psi (\boldsymbol{\varepsilon})}{\partial \boldsymbol{\varepsilon}}$ }is the Cauchy stress tensor. The regularization length $\ell>0$ is the length scale associated with the phase-field regularization of the fracture surface, and its value is important and necessary for numerical simulations to capture the crack propagation.
The element size $h$ of the finite element method should be smaller than $\ell$, and the parameter $\eta_{\ell}$ is set to a small value of the order of $o(\ell)$, see Figure \ref{body_1}.

The isotropic model in \cite{AMBATI_2015} depends on the history variable $H$ defined as:
\begin{equation}
H = 
\begin{cases}
\psi(\varepsilon) & \psi(\varepsilon) < H_n \\
H_n & \text{otherwise},
\end{cases}
\end{equation}
where $H_n$ is the strain energy computed at load step $n$. 

A modification of the phase-field method known as hybrid phase field method is analyzed in \cite{AMBATI_2015}, this model
is given by,
\begin{equation}
  \label{ec3}
 -G_c l_0 \nabla ^2 \phi + \left[ \frac{G_c}{l_o} + 2 H^+ \right] \phi = 2H^+, \quad \text{in } \quad \Omega,
\end{equation}
where $H^+ = max_{\tau \in [0,t]}\psi^+ \left( \varepsilon (x, \tau) \right)$.

To prevent crack face interpenetration, the model should satisfy the
constraint,

\begin{equation}
 \forall x: \psi^+ < \psi^- \Rightarrow \phi:=0,
\end{equation}
and 
\begin{equation}
 \psi^{\pm}(\varepsilon) = \frac12 \lambda \langle tr(\varepsilon) \rangle^2_{\pm} + \mu tr(\varepsilon^2_{\pm})
\end{equation}
with 
\begin{equation}
\varepsilon_{\pm} = \sum^3_{I = 1} \langle \varepsilon_I \rangle_{\pm}{\bf{n_I}}\otimes {\bf{n_I}}, \quad \text{and} \quad \varepsilon =  \sum^3_{I = 1} \langle \varepsilon_I \rangle \bf{n_I}\otimes \bf{n_I} ,
\end{equation}
where $\langle x \rangle_{+}:=(|x|+x)/2$ and $\langle x \rangle_{-}:=(|x|-x) / 2$, and $\varepsilon_I$ and $\bf{n_I}$ are the principle strains and principle strain directions, respectively.  

This model has the property that the linear momentum equation is retained as in the isotropic model and the evolution of the hybrid phase field variable is controlled by the tensile elastic energy $\psi^+$, ensuring that the crack surfaces do not grow in the compression regions, avoiding imaginary cracks. This hybrid formulation already reduces computational complexity and execution times \cite{HIRSHIKESH_2019}.

\subsection{The variational model}\label{sec_wvm}

The variational model considers a linear-elastic body with a crack $\Gamma_C$ in the domain $\Omega \subset \mathbb{R}^n$, where $n=2,3$.  Let $\boldsymbol{u}: \Omega \rightarrow \mathbb{R}^n$ be the displacement field. The general minimization problem considers the functional of the total energy of the body expressed by the sum of the bulk elastic energy and the fracture surface energy introduced by \cite{TANNE_2018,FRANCFORT_1998},

\begin{equation}\label{eq:fracfort}
%\left(u_i, \Gamma_i\right)=\underset{\substack{u=\bar{u}_i \text { on } \Gamma_D \Omega \\ \Gamma_C \supset \Gamma_{i-1}}}
\min {E}(\boldsymbol{u}, \Gamma_C):=\int_{\Omega \backslash \Gamma_C} \psi(\boldsymbol{\varepsilon}) d\Omega + G_c \int_{\Gamma_C}  dS.
\end{equation}
 The body is subjected to boundary conditions: tractions, $\bar{\tau}$, applied along the Neumann boundary $\Gamma_{N}$ and displacements, $\overline{\boldsymbol{u}}$, applied along the Dirichlet boundary $\Gamma_{D}$, see Figure \ref{body_1}. 
The fracture propagation is determined by minimizing of the energy functional.

A regularized formulation of the variational model \eqref{eq:fracfort} have their origins in the research of Bourdin {\it et al.}~\cite{BOURDIN_2000}.  These approaches are based on a regularization of the Mumford-Shah problem in image processing~\cite{AMBROSIO_1990,AMBROSIO_1992,BRAIDES_1998}. To enable numerical simulations, this regularized formulation introduces the scalar damage variable $\phi: \Omega \rightarrow [0, 1]$, describing the trajectory of the crack on $\Omega$.
The boundary value problem for a brittle fracture becomes to: find $(\boldsymbol{u}, \phi): \Omega
\rightarrow \mathbb{R}^d\times[0,1]$ satisfying the minimization
problem, 
\begin{equation}
  \label{eq:minimization}
 \min E_{\ell}(\boldsymbol{u}, \phi)
\end{equation}
where,
{\begin{equation}
 E_{\ell}(\boldsymbol{u}, \phi)=\int_{\Omega} \left[a(\phi)+\eta_{\ell}\right]\psi(\boldsymbol{\varepsilon}) d \Omega+\frac{1}{4 c_{w}} \int_{\Omega}G_{c}\left(\frac{w(\phi)}{\ell}+\ell|\nabla \phi|^{2}\right) d\Omega .
\end{equation}}
The symmetrized gradient of $\boldsymbol{u}$ defined as ${\varepsilon}(\boldsymbol{u}):=\frac{1}{2}\left(\nabla \boldsymbol{u}+\nabla^{T}
\boldsymbol{u}\right)$  denotes the strain tensor.  The elastic energy density function described in terms of the
elastic Lamm\'e constants is $\psi(\boldsymbol{\varepsilon}) = \frac{1}{2}A\varepsilon(\boldsymbol{u}):\varepsilon(\boldsymbol{u}) = \frac{1}{2} \lambda tr^2(\varepsilon(\boldsymbol{u})) + \mu tr(\varepsilon^2(\boldsymbol{u}))$, with $\lambda>0$ and $\mu>0$.
$a(\phi)$ and $w(\phi)$ are
continuous monotonic functions such that $a(0)=1$, $a(1)=0$, $w(0)=0,$ and
$w(1)=1$,  and
$c_{w}:=\int_{0}^{1} \sqrt{w(s)} d s$ is a normalization term. 

The crack is described by the smooth transition function $\phi$, and must satisfy the following conditions ~\cite{AMBATI_2015,MSEKH_2015,MIEHE_2010a}
\begin{itemize}
 \item is a symmetric function with respect to the crack,
 \item is a monotonic increasing function on time,
 \item the maximum value of $\phi$ on the crack is one and decays to zero as it moves away from the fracture.
\end{itemize}

The most common variational mathematical model in the literature
\cite{BLEYER_2018,TANNE_2018} employs $a(\phi) = (1-\phi)^2$ and $w(\phi) =
\phi^2$ and the functional becomes,
{\begin{equation}\label{eq21}
 E_{\ell}(\boldsymbol{u}, \phi)=\int_{\Omega} \left[(1-\phi)^{2}+\eta_{\ell} \right ]\psi(\boldsymbol{\varepsilon}) d \Omega +\frac{1}{2} \int_{\Omega}G_{c} \left( \frac{\phi^{2}}{\ell}+\ell|\nabla \phi|^{2} \right) d \Omega .
\end{equation}}

This model ensures the irreversibility of the crack phase-field evolution considering the condition $\phi \geq \phi_{n-1}$ in
$\Omega$, for a given loading step $n \geq 1$, details in \cite{MIEHE_2010,BLEYER_2018}. In \cite{BOURDIN_2008}, the author proves the $\Gamma$-converge of the regularized formulation \eqref{eq21} to the original formulation \eqref{eq:fracfort} as $\ell \rightarrow 0$. 

%We propose the weighted-variational model given by,
%\begin{equation}\label{eq:wv}
% E^*_{\ell}(\boldsymbol{u}, \phi) = \xi E_{\ell}(\boldsymbol{u}, \phi),
% \end{equation}
%where $\xi$ is a real constant, as a surrogate model for the hybrid phase field method.

\subsection{The weighted-variational approach as a surrogate mathematical model}

In this paper, following the idea of multi-fidelity, we propose a weighted-variational model depending of $\Delta_u, \ell, h$, as a surrogate model for the hybrid phase field method given by,
\begin{equation}\label{eq:wv}
 E^*_{ \Delta_u, \ell, h}(\boldsymbol{u}, \phi) = \xi E_{\ell}(\boldsymbol{u}, \phi:\Delta_u, \ell, h),
 \end{equation}
 where $\xi$ is a scale multiplicative
correction parameter given by,
$$\xi = \frac{\max \{\text{Reference displacement-force}\}}{\max \{\text{Surrogate displacement-force} \}},$$
similar scale parameters for surrogate models are found in \cite{fernandez2023review}. Then, our new boundary value
problem for brittle fracture is: find $(\boldsymbol{u}, \phi): \Omega
\rightarrow \mathbb{R}^d\times[0,1]$ satisfying the minimization
problem,
\begin{equation}
  \label{eq:minimization2}
 \min  E^*_{ \Delta_u, \ell, h}(\boldsymbol{u}, \phi)
\end{equation}
The idea behind this proposed surrogate model is to induce some leverage in the functional $E_{\ell}(\boldsymbol{u}, \phi)$ by calibrating $\Delta_u, \ell,$ and $ h$ to obtain approximate solutions to the original variational problem.  Note that the original expression in \eqref{eq21} is itself a simplified approach to reality and is not intended to be a solution of the continuous-time model in (\ref{eq:fracfort}).  In addition, the original variational problem must be carefully calibrated to correctly reproduce the crack evolution \cite{MIEHE_2010,AMBATI_2015}
Indeed, the introduction of $\xi$ allows a better calibration of the surrogate model and, as will be seen in the examples, the same parameter calibration process is required as for other benchmarks, including the hybrid phase model we use here.  \cite{HIRSHIKESH_2019}. The aim of this study is to approximate the profiles of displacement-force curves using the surrogate model. This will be done by comparing the curves reported in the literature with those obtained from the hybrid-phase model presented in this paper. A process is envisaged in which a few standard runs are performed with the expensive hybrid phase field model to find an adequate calibration, and the corresponding parameters are in turn used in our proposed weighted variational surrogate.  The calibration process will be explained in section~\ref{sec:calibration}, but for now we would like to mention that no additional (``judicious'') calibration is needed for this surrogate, thus making it a valid option \cite{Asher_2015}.

\section{FEniCS implementation}\label{sec:4}

In this section, we describe the numerical methodology for finding solutions to the deterministic problems~\eqref{ec3} and~\eqref{eq:wv}. In both cases we use FEniCS, an open source finite element software \cite{AlnaesEtal2015}, to obtain numerical simulations. 
As mentioned in \cite{BOURDIN_2000}, $\ell$ must be small enough to prevent the underestimated bulk energy from causing a softening effect and large enough compared to the discretization size $h$ near the cracks to avoid a surface energy overestimation. To satisfy this restriction the relation $\ell = h/k$ is suggested and the mesh size is restricted by $h=\ell/2$ \cite{MIEHE_2010,Shen_2018}, or more strictly  $l = O(h)$ \cite{MIEHE_2010}. In most cases, as long as the relationship is satisfied near the fracture, the accuracy can be ensured without over-resolving the problem \cite{Shen_2018}. Different values of $k$ are applied for fracture simulations in the literature, $k=1$ \cite{Cazes_2015}, $k=2$ \cite{HIRSHIKESH_2019}, and $k>2$ \cite{MIEHE_2010}.  A more in depth explanation of the relationship between $\ell$ and $h$ is presented in \cite{MIEHE_2010}. A study regarding the impact of the parameter $\Delta_u$ in the resulting simulations is reported in \cite{Singh_2016}. 

\begin{algorithm}
\caption{Staggered Scheme} \label{algoritmo2}
\hspace*{\algorithmicindent} {{\bf Input:} Initial parameters $ h, \ell, \Delta_u$. Gaussian random field $G_c$. Set $\left(\mathbf{u}^{0}, \phi^{0}\right)(\mathbf{x})=(\mathbf{0}, 0)$, $\forall \mathbf{x}\in\Omega$.}\\
\hspace*{\algorithmicindent} {{\bf Output:} Calibrated parameters $ h, \ell, \Delta_u$, approximated {$\mathbf{u}$} and $\phi$.}\\
\hspace*{\algorithmicindent} {{\bf Do} until {\bf calibration} of $h, \ell, \Delta_u$ the following steps:}
\begin{algorithmic}[1]
%Entrada: NÃºmero entero $geq2$\
\STATE { { Given } $\left(\mathbf{u}^n, \phi^{n}\right)$ at the previous loading step. }\\
\STATE { Set $\left[\phi^{n+1}\right]^{0}=\phi^{n}$ for $n>0$}\\
\WHILE {not converged}
\STATE { Compute $\left[\mathbf{u}^{n+1}\right]^{i+1}$ by solving the equation
$$
\nabla \cdot \boldsymbol{\sigma}\left(\left[\mathbf{u}^{n+1}\right]^{i+1},\left[\phi^{n+1}\right]^{i}\right)=0 \text { in } \Omega
$$ with $\text { boundary conditions } \left[\mathbf{u}^{n+1}\right]^{i+1}=\mathbf{u}_{D}^{n+1} \text { on } \Gamma_{D} \text {, and } \boldsymbol{\sigma} \cdot \mathbf{n}=\mathbf{t}^{n+1} \text { on } \Gamma_{N}.$}\\
\STATE{Compute the history field $\left[\mathcal{H}^{n+1}\right]^{i+1}=\max \left(\mathcal{H}^{n},\left[\Psi_{0}^{n+1}\right]^{i+1}\right)$.} \\
\STATE{Compute $\left[\phi^{n+1}\right]^{i+1}$ by solving in $\Omega$,
{
$$ -G_c \ell \nabla ^2 \left[\phi^{n+1}\right]^{i+1} + \left[ \frac{G_c}{\ell} + 2 \left[\mathcal{H}^{n+1}\right]^{i+1} \right] \left[\phi^{n+1}\right]^{i+1} = 2\left[\mathcal{H}^{n+1}\right]^{i+1}, 
$$
}
with boundary condition $\left(\nabla\left[\phi^{n+1}\right]^{i+1}\right) \cdot \mathbf{n}=0$ on $\partial \Omega$}
\ENDWHILE
\STATE { End}\\
\end{algorithmic}
\end{algorithm}

%In our study we calibrate three main parameters in numerical simulations for the hybrid phase field $h,\ell,\Delta_u$, \ac Luis: esto no se entiende, >$\xi$ no se calibra aqui sino en el weighted-variational?   and we include the parameter $\xi$ for the proposed weighted-variational model. ??? .  En ese caso, yo no diria nada aqui \ca

\subsection{Numerical methodology for the hybrid phase-field approach}\label{sec:hfpm}

The strategy to solve the hybrid phase-field method is the staggered scheme.  This algorithm computes the displacement field $\mathbf{u}$ and the damage field $\phi$ alternately until convergence at each load steps $\Delta u$ is achieved, see Algorithm \ref{algoritmo2}. This strategy iterates until a stop criterion for convergence is reached.  The alternatively is to take sufficiently small increments $\Delta u$ \cite{MUIXI_2020} to obtain a correct approximation of the crack, thus incurring in high computational costs. An analysis of the  iterations in the staggered scheme is presented in \cite{AMBATI_2015}. We calibrate the parameters $\ell, h, \Delta_u$ to obtain similar simulations to the those reported in the literature \cite{HIRSHIKESH_2019,AMBATI_2015,MUIXI_2020,HUYNH_2019}

 Generally, the equations of the hybrid phase field method \eqref{ec3} in the staggered scheme are solved  using the finite element method \cite{HIRSHIKESH_2019,HUYNH_2019}.

\subsection{Numerical methodology for the weighted-variational approach}

In order to solve the weighted-variational problem~\eqref{eq:wv}, an iterative algorithm known as alternate minimization \cite{BOURDIN_2000,MARIGO_2016} is presented in Algorithm \ref{algoritmo1}.  We assume a uniform time step $\Delta t$. This numerical technique uses the displacement $\mathbf{u}_{i-1}$ and the damage field $ \phi_{i-1}$ at time step $t_{i-1}$.  In turn, the solution for $t_{i}$ is obtained by solving the following minimization problem
\begin{equation}\label{ec:min}
 \min \left\{ E_{\ell,t_{i}}(\mathbf{u}, \phi)|  \mathbf{u} \in \mathcal{C}_{t_{i}},  \phi \in \mathcal{D}_{t_{i}}\right\}
 \end{equation}
where $\mathcal{C}_{t_{i}}(\Omega)= \left\{u \in H^{1}\left(\Omega \right)| u=\mathbf{u}_{i-1}\right.$ on $\left.\partial_{\mathbf{u}_{i-1}} \Omega\right\}$, and $\mathcal{D}_{t_{i}}=\left\{\phi \in H^{1}(\Omega): \phi(x) \geq \phi_{i-1}\right. \text{a.e. in }\Omega, \phi=\phi_{i-1}$ on $\partial_{\phi_{i-1}} \Omega\}$. This model should satisfy the irreversibility of damage, which in its discrete version is
$\phi(x) \geq \phi_{i-1}$. Following previous analysis in \cite{BOURDIN_2000} and \cite{AMBROSIO_1992}, for  $a(\alpha)=(1-\alpha)^{2}+\eta_{\ell}$ and $w(\alpha)=\alpha^{2},$ with $\eta_{\ell}=o(\ell)$, it is possible to show through asymptotic methods the called $\Gamma$-convergence, this is,  if $ E_{\ell,t_{i}}$ converges to $E$ as $\ell \rightarrow 0$, then the global minimizers of $ E_{\ell,t_{i}}$ converge to that of $E$, more details in \cite{TANNE_2018,FRANCFORT_1998,GIACOMINI_2005}.

%Let us consider the general minimization problem introduced from \cite{TANNE_2018,FRANCFORT_1998},
%\begin{equation}
%\left(u_i, \Gamma_i\right)=\underset{\substack{u=\bar{u}_i \text { on } \partial_D \Omega \\ \Gamma \supset \Gamma_{i-1}}}{\operatorname{min}} {E}(u, \Gamma):=\int_{\Omega \backslash \Gamma} \psi(\boldsymbol{\varepsilon}) d\Omega + G_c \int_{\Gamma}  dS,
%\end{equation}
%the  solutions of the global minimization problem \eqref{ec:min} tend to the solution of the global minimization \lb functional \bl \eqref{eq21} as $\ell \rightarrow 0$, .

For this proposed model, we use Algorithm \ref{algoritmo1} to calibrate the four parameters $ h,\ell, \Delta_u, \xi$ that would reproduce similar profiles of displacement-force curves as those reported in the literature \cite{MIEHE_2010,AMBATI_2015}.

\begin{algorithm}
\caption{Alternate minimization}\label{algoritmo1}
\hspace*{\algorithmicindent} {{\bf Input:} Initial parameters $ h, \ell, \Delta_u,\xi$. Gaussian random field $G_c$. Set $\left(\mathbf{u}^{0}, \phi^{0}\right)(\mathbf{x})=(\mathbf{0}, 0)$, $\forall \mathbf{x}\in\Omega$.}\\
\hspace*{\algorithmicindent} {{\bf Output:} Calibrated parameters $ h, \ell, \Delta_u, \xi$, approximated {$\mathbf{u}$} and $\phi$.}\\
\hspace*{\algorithmicindent} {{\bf Do} until {\bf calibration} of $h, \ell, \Delta_u, \xi$, the following steps:}
\begin{algorithmic}[1]
\STATE { { Given the state } $\left(\mathbf{u}_{i-1}, \phi_{i-1}\right)$ at the loading step $t_{i-1}$. }\\
\STATE {{ Set } $\left(\mathbf{u}^{(0)}, \phi^{(0)}\right):=\left(\mathbf{u}_{i-1}, \phi_{i-1}\right)$}\\
\WHILE {not converged}
\STATE { { Find } $\mathbf{u}^{(s)}:=\underset{\mathbf{u} \in \mathcal{C}_{t_{i}}}\arg \min  {E_{\ell}^*}_{t_{i}}\left(\mathbf{u}, \phi^{(s-1)}\right)$ }\\
\STATE{{ Find } $\phi^{(s)}:=\underset{\phi \in \mathcal{D}_{t_{i}}}\arg \min  {E_{\ell}^*}_{t_{i}}\left(\mathbf{u}^{(s)}, \phi\right)$} \\
\ENDWHILE
\STATE {{ Set } $\left(\mathbf{u}_{i}, \phi_{i}\right)=\left(\mathbf{u}^{(s)}, \phi^{(s)}\right) $}\\
\STATE { End}\\
\end{algorithmic}
\end{algorithm}

\section{Random energy release rate $G_c$}\label{sec:3}

The behavior of materials under stress is primarily determined by their microstructure. For many materials, the microstructure can rarely be characterized from a purely deterministic point of view since the size and spatial distribution of the defects that make up the microstructure tend to be random, such as glass, ceramics, concrete, and some alloys. Many material properties of brittle materials, such as fracture strength or fracture energy, must be modeled as random. However, material properties are often spatially dependent and can be effectively treated using random fields~\cite{YANG_2008,YANG_2009,Jeulin_1994}. In the context of brittle fracture, it is clear that a fracture path is strongly influenced by the inherent material randomness.

We propose to model the randomness associated with brittle fracture in terms of Gaussian random fields; in particular, we assume that the energy release rate $G_c$ is a stationary Gaussian random field. 

A Gaussian random field is defined as a random function $f: \Omega \rightarrow \R$ such that any finite collection of random variables $f(\boldsymbol{x}_1), \ldots,f(\boldsymbol{x}_n)$ at points $\boldsymbol{x}_1,\ldots,\boldsymbol{x}_n \in \Omega$ have a multidimensional Gaussian distribution \cite{YANG_2008,Sarkka_2013}. A Gaussian field is fully characterised in terms of its mean $m(\boldsymbol{x})$ and covariance function $k\left(\boldsymbol{x}, \boldsymbol{x}^{\prime}\right)$, 
\[
\begin{aligned}
m(\boldsymbol{x}) &=\mathrm{E}[f(\boldsymbol{x})] \\
k\left(\boldsymbol{x}, \boldsymbol{x}^{\prime}\right) &=\mathrm{E}\left[(f(\boldsymbol{x})-m(\boldsymbol{x}))\left(f\left(\boldsymbol{x}^{\prime}\right)-m\left(\boldsymbol{x}^{\prime}\right)\right)\right] .
\end{aligned}
\]
The joint distribution of any finite collection of random variables $f\left(\boldsymbol{x}_{1}\right), \ldots, f\left(\boldsymbol{x}_{n}\right)$ is then multidimensional Gaussian,
\[
\left(\begin{array}{c}
f\left(\boldsymbol{x}_{1}\right) \\
\vdots \\
f\left(\boldsymbol{x}_{n}\right)
\end{array}\right) \sim \mathrm{N}\left(\left(\begin{array}{c}
m\left(\boldsymbol{x}_{1}\right) \\
\vdots \\
m\left(\boldsymbol{x}_{n}\right)
\end{array}\right),\left(\begin{array}{ccc}
k\left(\boldsymbol{x}_{1}, \boldsymbol{x}_{1}\right) & \cdots & k\left(\boldsymbol{x}_{1}, \boldsymbol{x}_{n}\right) \\
\vdots & \ddots & \\
k\left(\boldsymbol{x}_{n}, \boldsymbol{x}_{1}\right) & & k\left(\boldsymbol{x}_{n}, \boldsymbol{x}_{n}\right)
\end{array}\right)\right) .
\]
A Gaussian process is said to be stationary if its mean is constant and the covariance function is of the form \cite{Sarkka_2013},
\[
k\left(\boldsymbol{x}, \boldsymbol{x}^{\prime}\right)=C\left(\boldsymbol{x}^{\prime}-\boldsymbol{x}\right)
\]
where $C(\boldsymbol{x})$ is another function, the stationary covariance function of the process.

\begin{figure*}[h!]
%\centering
\begin{subfigure}[b]{.6\linewidth}
\includegraphics[width=9.5cm]{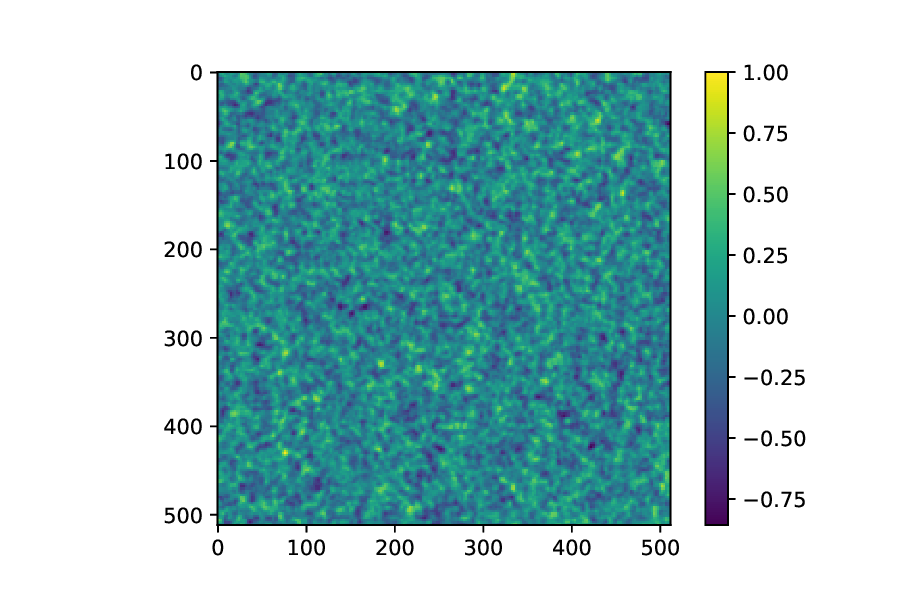}
\caption{}
\end{subfigure}
\begin{subfigure}[b]{0.3\linewidth}
\includegraphics[width=6.4cm]{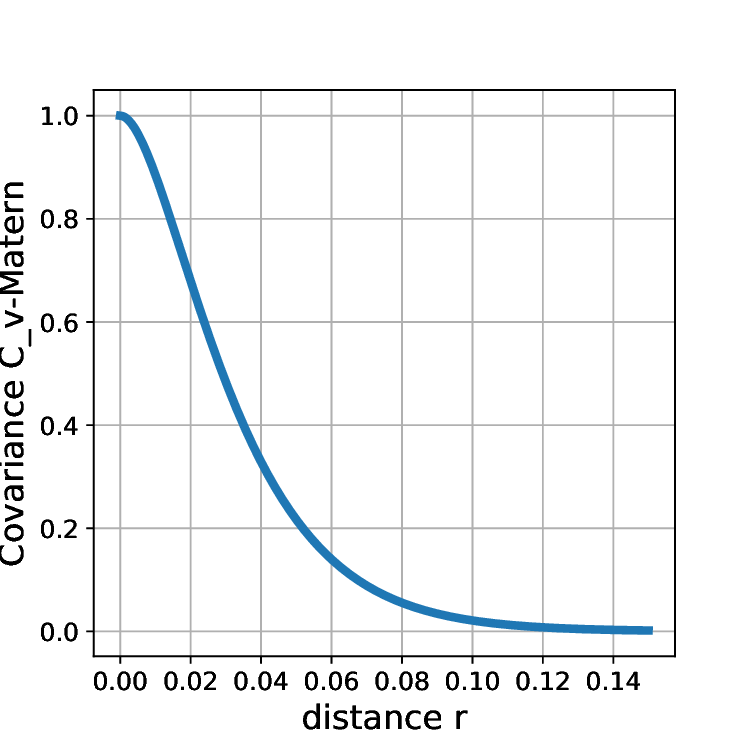}
\caption{}
\end{subfigure}
\caption{a) A 2D Gaussian random field with Mat\'ern covariance;  b) Mat\'ern covariance, with $\nu=3/2$, as a function of distance $r$.}\label{im1}
\end{figure*}

In order to obtain the heterogeneity on $G_c$, we use a Gaussian random field that is generated by the multidimensional Mat\'ern covariance function,
\begin{equation}\label{eqn:matern}
C(r)=\sigma^{2} \frac{2^{1-v}}{\Gamma(\nu)}\left(\sqrt{2 \nu} \frac{r}{l}\right)^{v} K_{v}\left(\sqrt{2 v} \frac{r}{l}\right)
\end{equation}
where $r=\left\|\boldsymbol{\xi}-\boldsymbol{\xi}^{\prime}\right\|$,  $ \boldsymbol{\xi}=\left(x_{1}, x_{2}, \ldots, x_{d-1}, t\right) \in \mathbb{R}^{d}$,  $\Gamma$ is the gamma function, $K_{\nu}$ is the modified Bessel function of the second kind, and $l$ and $\nu$ are positive parameters of the covariance. 

A non-complex way to obtain a random field follows the classical Wiener-Khinchin theorem, which states that the stationary covariance function of a process can be expressed as the inverse Fourier transform of the spectral density~\cite{Sarkka_2013},
\begin{equation}
C(t)=\mathcal{F}^{-1}[S(\omega)]=\frac{1}{2 \pi} \int S(\omega) \exp (i \omega t) \mathrm{d} \omega.
\end{equation}

Applying this idea, we can recover a Gaussian random field with Mat\'ern covariance using the corresponding spectral density for the Mat\'ern covariance matrix,
\begin{equation}
S\left(\omega_{r}\right)=S\left(\omega_{x}, \omega_{t}\right) \propto \frac{1}{\left(\theta^{2}+\left\|\omega_{x}\right\|^{2}+\omega_{t}^{2}\right)^{v+d / 2}},
\end{equation}
where $\theta=\sqrt{2 v} / l$. This allows the process to be easily simulated, even for very large grids \cite{Sarkka_2013}.
The use of the Mat\'ern covariance function is justified (especially for the particular case of $\nu=3/2$, in which case (\eqref{eqn:matern}) has an analytic form), since its Fourier transforms have been computed, making it suitable for spectral analysis, as above \cite{Sarkka_2013}.  The Mat\'ern autocovariance, with $\nu=3/2$, decays mostly exponentially but, crucially, is smooth at zero, as opposed to a simple exponential \cite{Ras_2006}.  The corresponding Gaussian process is smooth, with a continuous derivative.
The correlation length is controlled by the scale parameter $l$.  Thus, a combination of mathematical convenience and parsimonious modelling flexibility is the only justification for the now widely used Mat\'ern covariance function. 

\section{Results and simulations}\label{sec:5}

\subsection{Calibration of the hybrid phase field model}

\begin{figure}[H]
    \begin{center} 
  \includegraphics[width=10 cm]{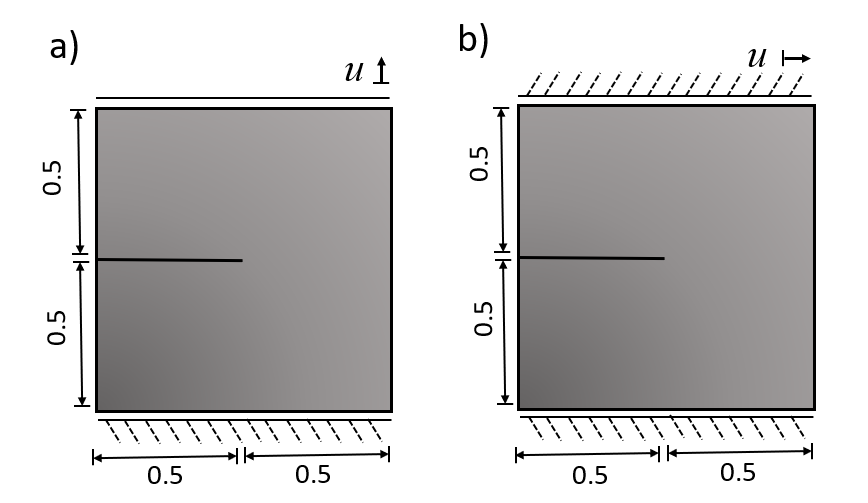}
\end{center} 
\caption{Plate with the initial crack: Geometry in mm, and boundary conditions for (a) Tension and (b) Shear test.}\label{figcf}
\end{figure}

To calibrate our algorithms, we consider a square plate with a straight edge crack (standard for calibrating algorithms, \cite{HIRSHIKESH_2019}). We use two types of test problems: first, the pure mode I fracture, the plate is subjected to a displacement in the y-direction at the top to simulate a brittle fracture, the geometry and boundary conditions are given in Figure \ref{figcf}(a). The expected fracture propagation is a crack in the direction of the initial fracture. The second test, shear mode II fracture, considers the same plate loaded in shear mode, see figure \ref{figcf}(b) to observe the boundary conditions, in this case the expected initial crack propagation is a curve in the lower right hand side of the specimen.
  
  \begin{table}[h]
\indexspace
\begin{center}
%\resizebox{9cm}{!}{ 
\begin{tabular}{cc}\hline\hline\ 
Parameter & Value \\
\hline
$\lambda$ & 121.15 kN/mm$^2$ \\
$\mu$ & 80.0 kN/mm$^2$ \\
$G_c$ &  2.7 \\
$h$ & 0.0055\\
$\ell$ & 0.011 \\
\hline
\end{tabular}
%}
\caption{Fixed parameters for simulations.}\label{t_mpt}
\end{center}
\end{table}
  
\begin{figure*}[h!]
\centering
\begin{subfigure}[b]{.45\linewidth}
\caption{}
\includegraphics[width=8cm]{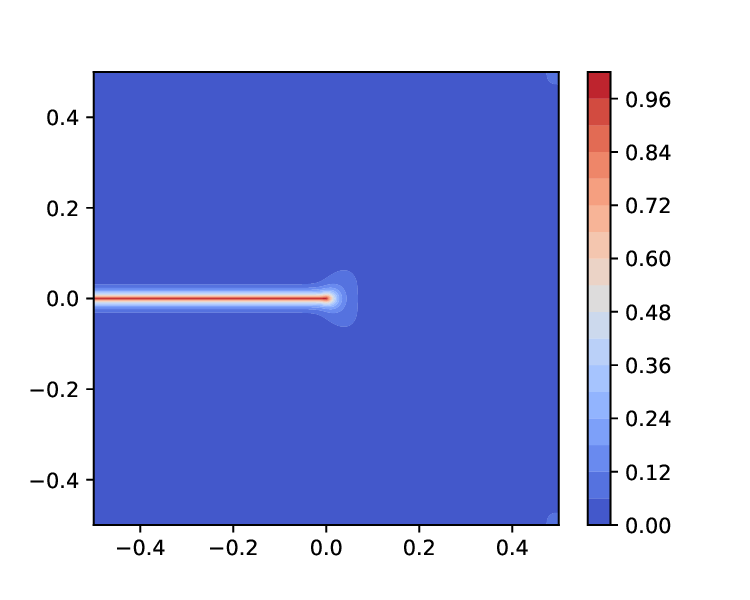}
\end{subfigure}
\begin{subfigure}[b]{0.45\linewidth}
\caption{}
\includegraphics[width=8cm]{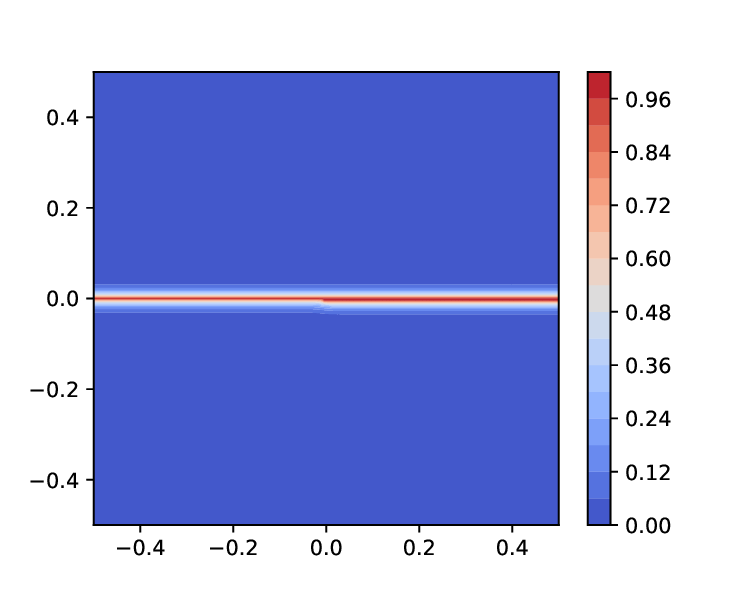}
\end{subfigure}
\caption{Status of the mode-I fracture of the hybrid phase field model at displacements of a) $5.3\times10^{-3}$ mm, and b) $5.4\times10^{-3}$ mm.}\label{frac1}
\end{figure*}

\begin{figure}[H]
\begin{center} 
  \includegraphics[width=8 cm]{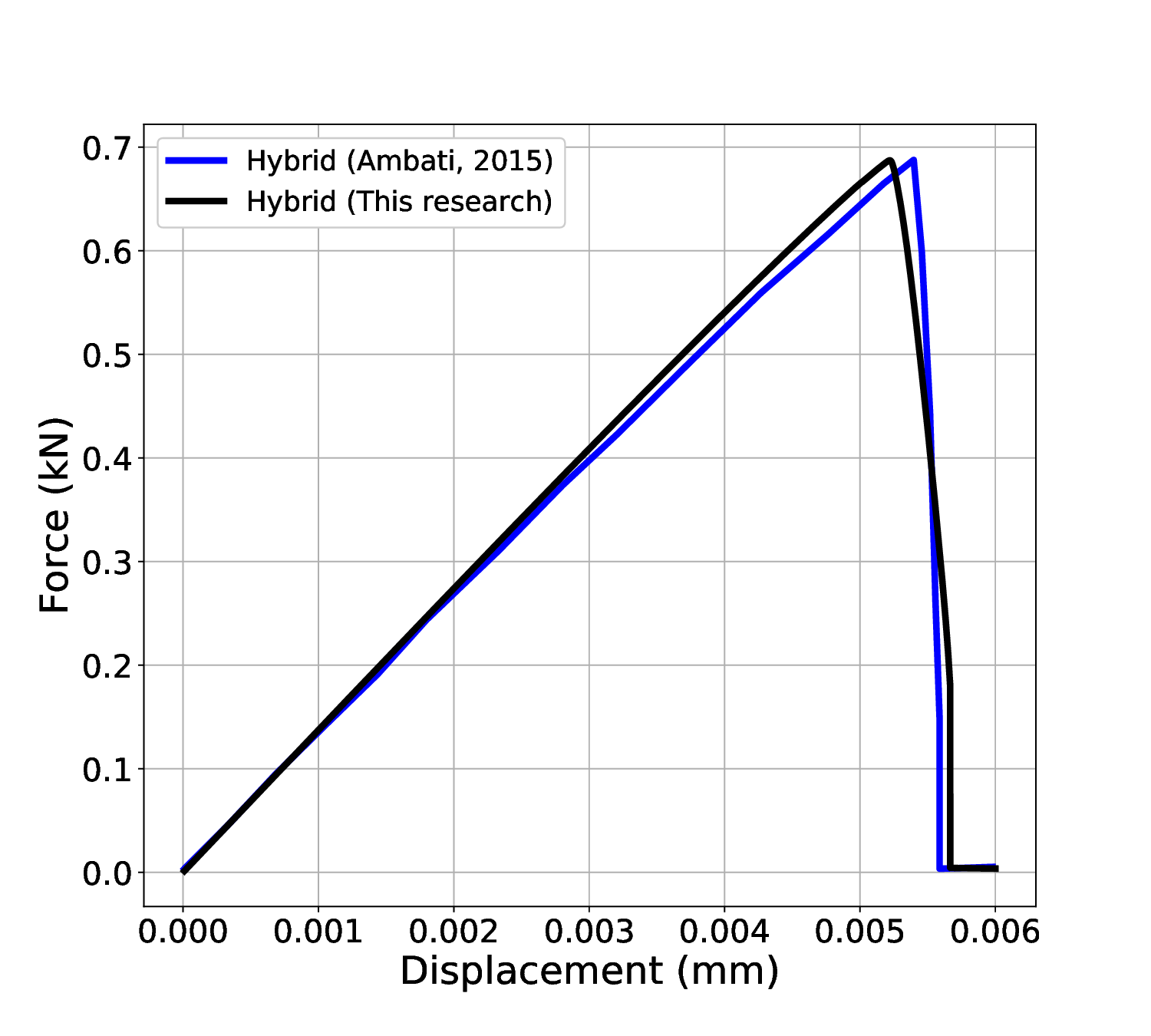}
\end{center}
\caption{Displacement-force of the hybrid phase field model and the reported in \cite{AMBATI_2015}. }
\label{frac_comp_}
\end{figure}

Simulations using the hybrid phase-field model were performed using the parameter set shown in table \ref{t_mpt}. For the mode I fracture we use 131,072 triangular elements, $\ell = 0.011$, and we applied displacement increments of $\Delta u = 1\times 10^{-5}$ mm up to $u = 5\times 10^{-3}$, and $\Delta u = 1\times 10^{-6}$ up to $u= 6\times 10^{-3}$. The numerical propagation of the initial crack is similar to the expected propagation, see Figure \ref{frac1}. The displacement-force plot is shown in Figure \ref{frac_comp_}, we observe the similar behaviour of the simulated profile and the profile reported in \cite{AMBATI_2015}.

In the second example, the shear mode II fracture, the same discretisation of the pure mode I fracture was used. The resulting propagation of the initial crack is a special case in which we study the parameter dependencies over simulations (mesh size $h$, $\ell$, $\Delta u$). It is well known that the solutions to brittle fracture problems are notoriously mesh-dependent, \cite{MIEHE_2010,HIRSHIKESH_2019,AMBATI_2015,HUYNH_2019}, each author of these papers fixing different values of mesh size $h$, $\ell$ and $\Delta u$ and obtaining different profiles that try to explain the behaviour of the uncertain crack. Different reported parameter configurations for numerical simulations are shown in the table \ref{tabla_sim}. For example, beak-like force-displacement profiles are found in~\cite{HIRSHIKESH_2019,AMBATI_2015,HUYNH_2019}, while mountain-like profiles are presented in~\cite{MIEHE_2010,WEINBERG_2017}.

\begin{figure}[H]
\begin{center} 
  \includegraphics[width=8 cm]{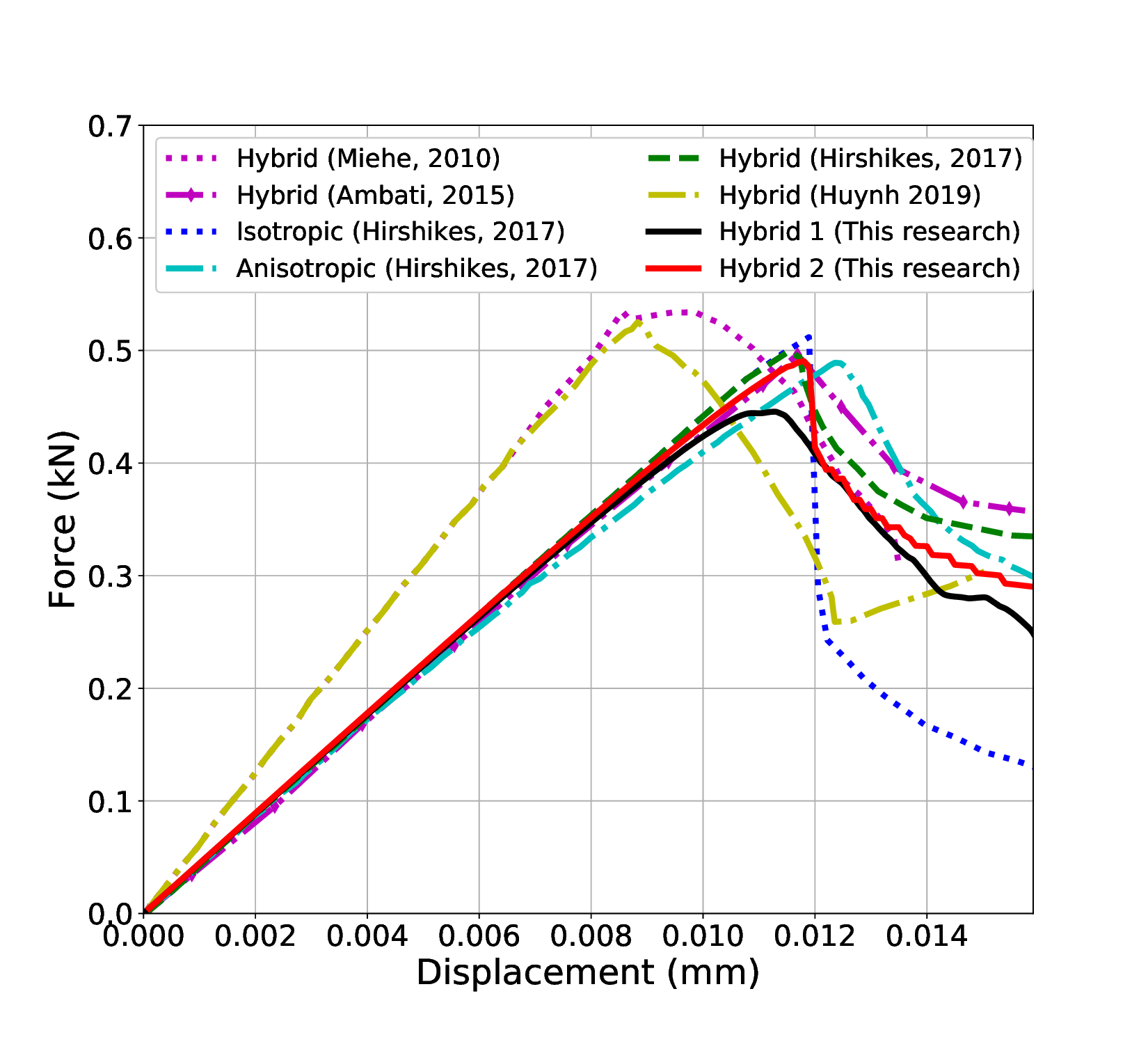}%{Images/comparative_lit.png}
\end{center}
\caption{Displacement-force curves for the shear mode-II fracture in this work compared with the reported in the literature, each configuration of mesh size $h$, $\ell$, and $\Delta u$ provides a characteristic profile (Table \ref{tabla_sim}).}\label{frac_comp}
\end{figure}

\begin{figure*}[h!]
\centering
\begin{subfigure}[b]{.45\linewidth}
\caption{}
\includegraphics[width=7cm]{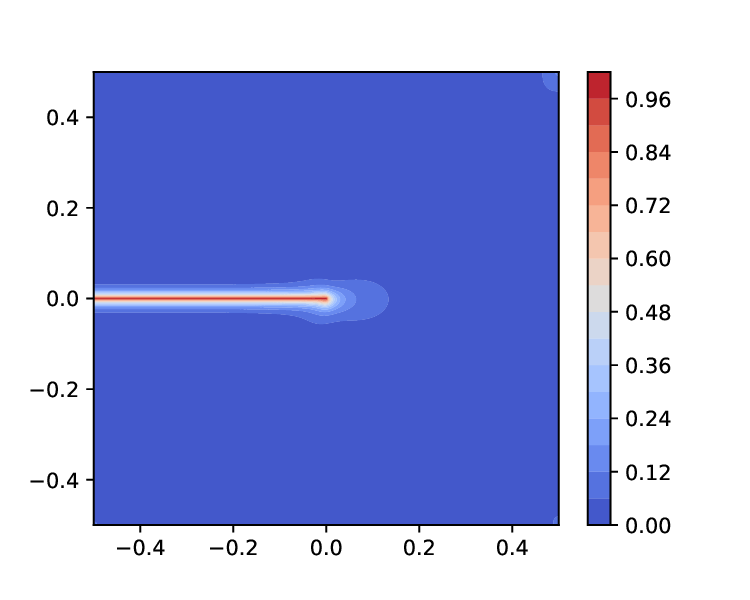}
\end{subfigure}
\begin{subfigure}[b]{0.45\linewidth}
\caption{}
\includegraphics[width=7cm]{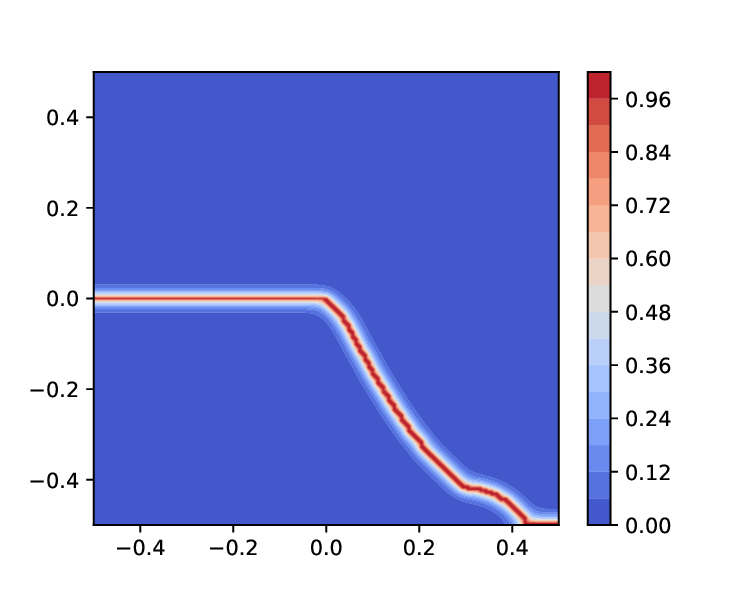}
\end{subfigure}
\caption{Status of the shear mode-II fracture with the hybrid P-F model at displacements of a) $5.3\times10^{-3}$ mm, and  b) $16\times10^{-3}$ mm.}\label{frac_shear}
\end{figure*}

We present our results for two different parameter configurations for the shear mode II fracture, see Table \ref{tabla_sim}. In our first simulation (Hybrid 1) we assume $\ell = 0.0087$ with a displacement increment of $\Delta u=1 \times 10^{-4}$ mm up to $u = 1.5\times ^{-2}$ mm. Here the force-displacement profile agrees well with the results in ~\cite{MIEHE_2010,WEINBERG_2017}. For our second simulation (Hybrid 2) we use $\ell = 0.0116$ with a displacement increment of $\Delta u=1 \times 10^{-5} \mathrm{mm}$, the resulting force-displacement profile contains a beak around the displacement of $0.011$~mm. Our simulation methodology is able to capture the essential features of brittle fracture (see Figure~\ref{frac_shear}). Figure \ref{frac_comp} shows the comparative profiles reported by other authors \cite{HIRSHIKESH_2019,AMBATI_2015}; we observe different profiles after $u = 0.01$ mm. 

  \begin{table}[h]
\indexspace
\begin{center}
\caption{Fixed parameters for simulations. $^a$ = triangular elements, $^b$ = quiadrilateral elements, $^* =$  refinement in zones where crack is expected to grow, $^{**}$=Authors uses the anisotropic model from Amor {\sl et al,} 2009 \cite{AMOR_2009}.}\label{tabla_sim}
\resizebox{9cm}{!}{ 
\begin{tabular}{p{115pt}p{60pt}p{38pt}p{39pt}p{30pt}p{38pt}}%{cccccc}
\hline\hline\ 
Researches & Model & FEM & $h$ (mm) & $\ell$  (mm) & $\Delta u$ (mm)\\
\hline
Miehe, 2010 \cite{MIEHE_2010} & Hybrid P-F & 30000$^a$ & $\approx 0.002$ & $0.015$ & $1 \times 10^{-5}$\\

Ambati, 2015 \cite{AMBATI_2015} & Hybrid P-F & 20592$^b$ & $0.001$ & $0.004$ & $1 \times 10^{-5}$\\ 

Hirshikesh, 2017 \cite{HIRSHIKESH_2019} & Isotropic  & 131072$^a$ & $0.0055$ &  $0.011$ & $1 \times 10^{-5}$\\

Hirshikesh, 2017 \cite{HIRSHIKESH_2019} & Hybrid P-F & 131072$^a$ & $0.0055$ &  $0.011$ & $1 \times 10^{-5}$\\

Hirshikesh${**}$, 2017 \cite{AMOR_2009,HIRSHIKESH_2019} & Anisotropic & 131072$^a$ & $0.0055$ &  $0.011$ & $1 \times 10^{-5}$\\

Huynh, 2019 \cite{HUYNH_2019} & Hybrid P-F& 25000$^{a,*}$  &  $\approx 0.004$  & $0.015$ & $1 \times 10^{-5}$ \\

This work & Hybrid 1 & 131072$^a$ & $0.0055$ & $0.0087$ & $1 \times 10^{-4}$ \\

This work & Hybrid 2 & 131072$^a$ & $0.0055$ & $0.01104 $ & $1 \times 10^{-5}$\\
\hline
\end{tabular}
}
\end{center}
\end{table}

This behaviour describes the shape, intensity and velocity of the crack generated in the specimen. The displacement-force curve shows a different profile when the mesh is refined over the region where the crack is expected \cite{HUYNH_2019}, and a maximum value around $u = 0.008$ mm with a mountain-like profile as reported in \cite{MIEHE_2010a}.

\subsection{The weighted-variational approach as a surrogate mathematical model}\label{sec:calibration}

In this section, we propose to use the weighted variational approach to construct a surrogate model of the hybrid phase-field method to reduce computational cost and complexity, as described in section \ref{sec_wvm}. %\ref{sec:hfpm}. 
Usually the computational cost of the hybrid phase-field method depends on the displacement increment $\Delta u$, the mesh size $h$ and $\ell$, this last parameter is usually taken to be $\ell = k h$.  The value $k=2$ has been suggested~\cite{MIEHE_2010,HIRSHIKESH_2019}. The computational cost increases when $\Delta u$ is small, e.g. for a maximum displacement of $0.6$ and an incremental displacement of $\Delta u = 1\times10^{-4}$ we need to solve the system \eqref{sis_1} 6,000 times.  If we change $\Delta u = 1\times10^{-5}$, 60,000 simulations should be performed.  In addition, the staggered scheme depends on the number of iterations to stabilise each solution according to the incremental load $\Delta u $, \cite{AMBATI_2015}.  On the other hand, if the mesh size $h$ of the discretisation of $\Omega$ decreases, the execution time increases.  

The weighted-variational approach should be able to accurately capture fracture, in principle, compared to the hybrid phase-field model, utilizing a much lower computational cost.  Thus, we anticipate that the weighted-variational model offers a promising alternative as a surrogate mathematical model for the hybrid phase field model.

\begin{figure*}[h!]
\centering
\begin{subfigure}[b]{.45\linewidth}
\caption{}
\includegraphics[width=7cm]{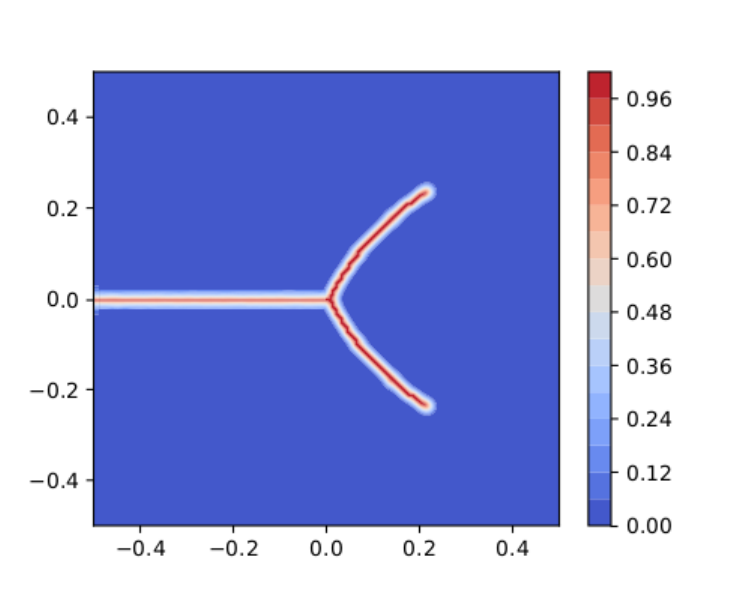}
\end{subfigure}
\begin{subfigure}[b]{0.45\linewidth}
\caption{}
\includegraphics[width=7cm]{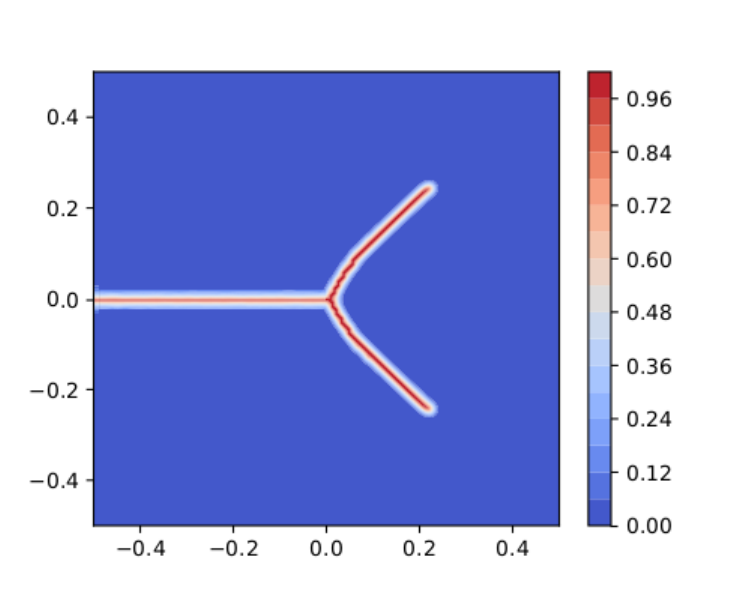}
\end{subfigure}
\caption{Comparative results of the W-V model with discretization of the displacement axis of:  a) 10 points, b) 50 points.}\label{sigma_111}
\end{figure*}

In order to examine our proposal, we compare the outcomes of the hybrid phase-field model and the weighted-variational model (W-V model 1) using $\xi = 1$ from Eq.~\eqref{eq:wv} for shear mode-II fractures. The body $\Omega$ has been discretized into 32,768 triangular components utilizing $\ell=0.016$. The applied displacements have been defined as $\Delta_u =t_M/N_{nodes}$, where $t_M$ stands for the maximum load and $N_{nodes}$ reflects the number of nodes.  We illustrate two simulations for shear fractures with $N_{nodes} = \{10, 20\}$ in figure \ref{sigma_111}, where we observe that the outcomes are quite similar. Both simulations exhibit identical crack trajectories.  

\begin{figure}[H]
    \begin{center} 
  \includegraphics[width=8 cm]{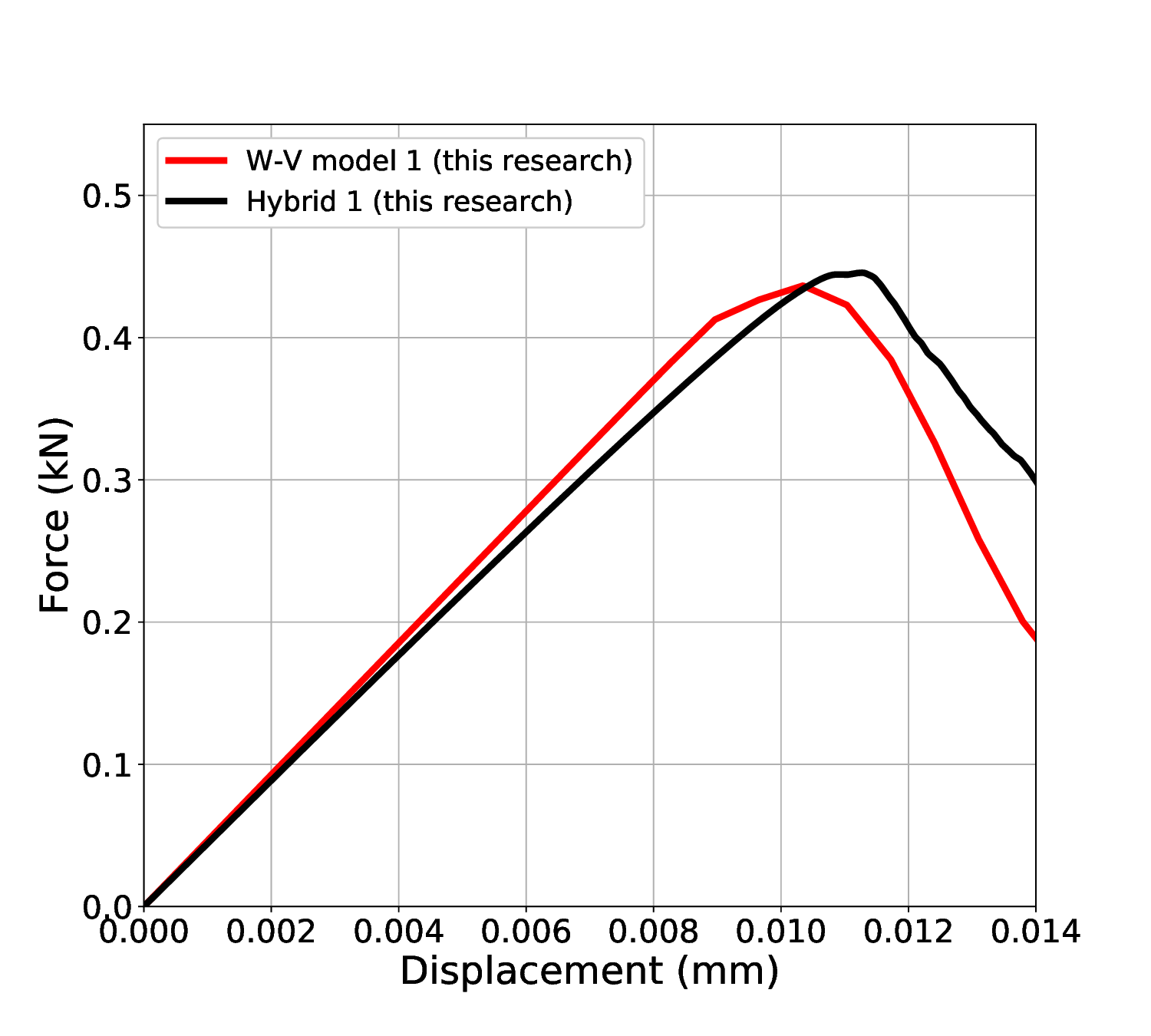}
\end{center} 
\caption{Comparative results of displacement-force obtained from the weighted-variational model using 20 points (blue line, $\approx$ 20 min.)
  and phase field method using 1024 points (red line, $\approx$ 8
  hrs.) .}\label{sigma_com1}
\end{figure}

The imaginary fracture discussed by other authors in previous studies is observed \cite{MIEHE_2010,HIRSHIKESH_2019}. However, the strength-displacement profiles are useful for describing the fracture's behaviour during displacement application. In this instance, the similar profiles indicate that the crack generated at around $0.010$ mm displacement, as shown in Figure \ref{sigma_com1}.

\begin{figure}[H]
%\begin{multicols}{2}
    \begin{center} 
  \includegraphics[width=8 cm]{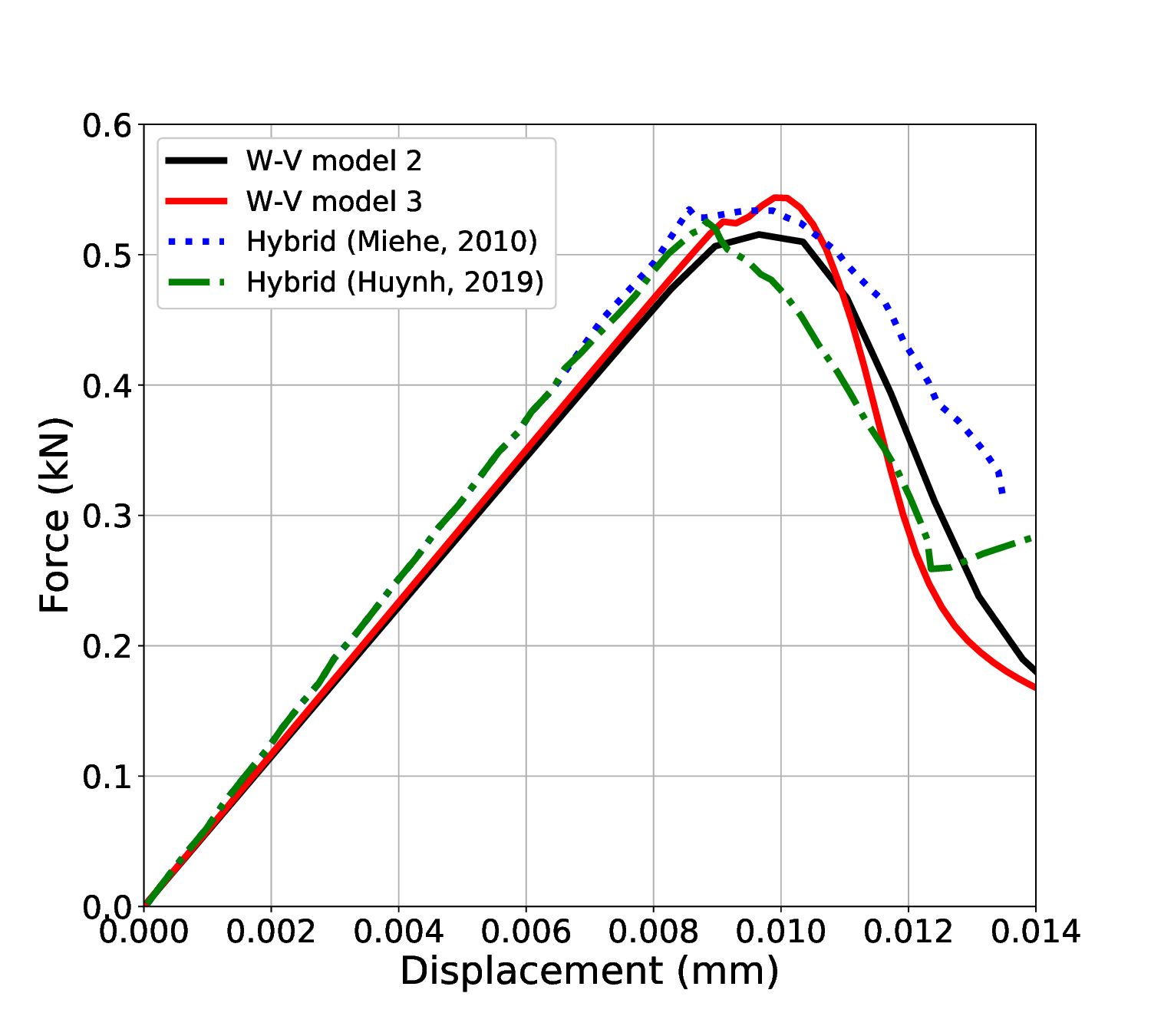}
\end{center} 
\caption{Results of the weighted-variational model with 30 nodes (black line) and 100 nodes (red line), compared with the reported in the literature.}\label{sigma_com2}
\end{figure}

We present the second set of simulations obtained from the Weighted-Variational Model 2 (W-V Model 2) and Weighted-Variational Model 3 (W-V Model 3).  In this instance, we have approximated the solutions previously reported in \cite{MIEHE_2010a} and \cite{HUYNH_2019}, see Table \ref{tabla_sim}. W-V Model 2 uses the same triangulation as V-W Model 1, having $N_{nodes} = 30$, and we calibrated it to $\xi = 1.28$. W-V Model 3 was employed with $N_{nodes} = 100$ and $\xi = 1.26$. This model causes the particular first beak around $u = 0.008$ mm.

The reported profiles in the literature and the profile obtained by the hybrid phase field model (Hybrid 1) are approximated correctly by our three parameter configurations (W-V model 1, 2, and 3), see Figures \ref{sigma_com1} and \ref{sigma_com2}.

\subsection{Gaussian random field $G_c$}

\begin{figure}[H]
%\begin{multicols}{2}
    \begin{center} 
  \includegraphics[width=8 cm]{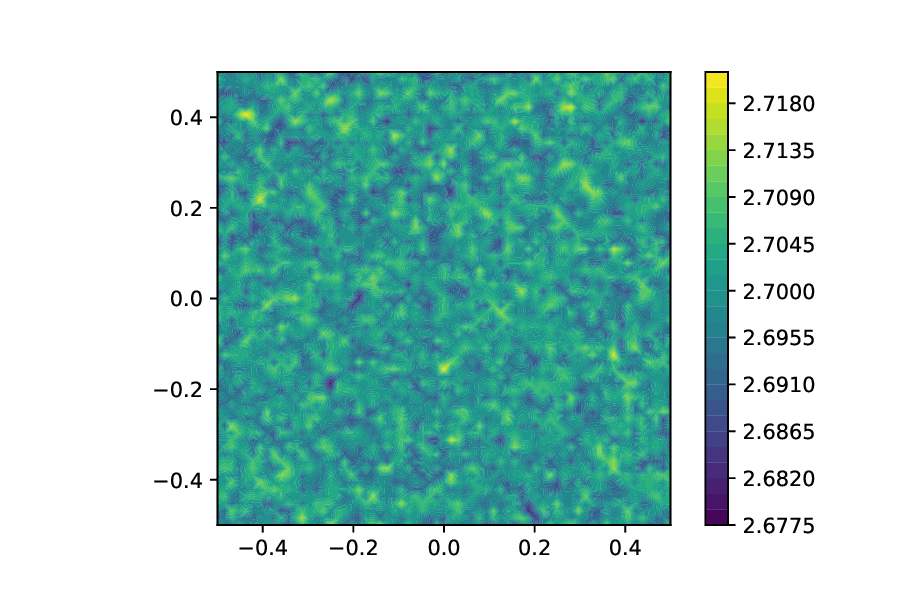}%{Images/cov_4.png}
\end{center} 
\caption{Example of a realization of $G_c$ modelled as a Gaussian random field, with Mat\'ern covariance.}\label{sigma_1_}
\end{figure}

This section considers a stationary Gaussian random field $G_c$ on $\Omega$ to explore how the randomness affects the solutions of $\boldsymbol{u}$ and $\phi$. Figure~\ref{sigma_1_} shows a sample of the Gaussian random field with Mat\'ern covariance for $G_c$. Table \ref{t_mpt} presents the parameters used for the simulations of a pure mode I fracture and shear mode II fracture, both using $l =$ 0.011. 
 Our models propagate the uncertainty in $G_c$ modeled by the later Gaussian process, to produce a distribution of fracture trajectories.  From a mathematical perspective, this is justifiable since the weak form assumes the applicability of $G_c \longmapsto (\boldsymbol{u},\phi)$, implicitly establishing the regularity and measurability of $(\boldsymbol{u},\phi)$ and properly defining a probability measure for them.

\begin{figure*}[h!]
\centering
\begin{subfigure}[b]{.45\linewidth}
\caption{}
\includegraphics[width=7cm]{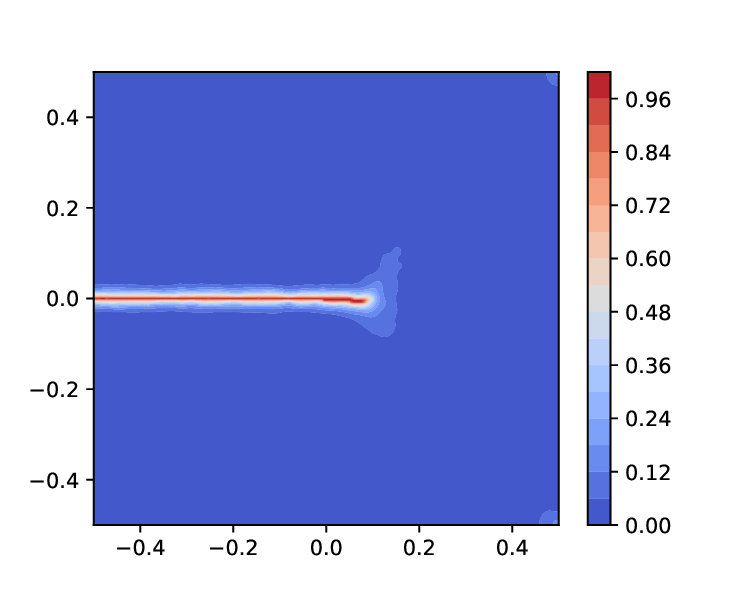}
\end{subfigure}
\begin{subfigure}[b]{0.45\linewidth}
\caption{}
\includegraphics[width=7cm]{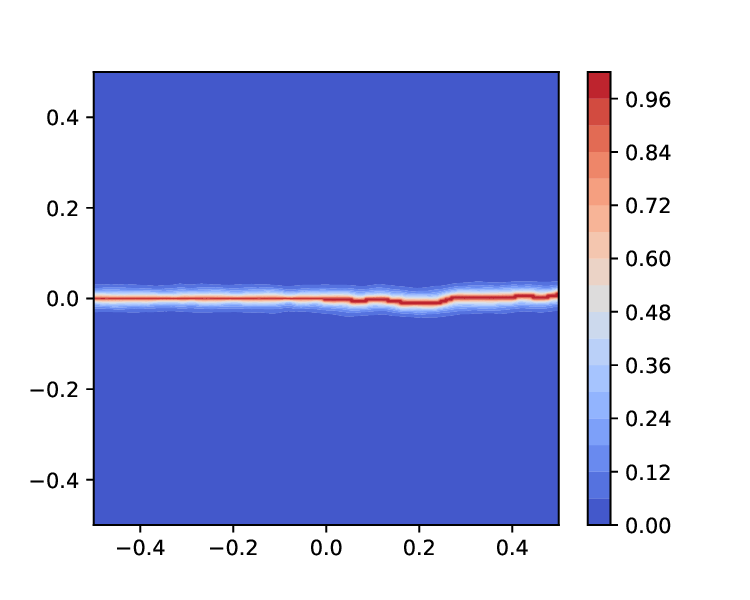}
\end{subfigure}
\caption{Trayectory of a mode-I fracture with random field $G_c$ at:  (a) $5.3\times10^{-3}$ mm; (b) $5.4\times10^{-3}$ mm.}\label{RM_SR1}
\end{figure*}

\begin{figure}[H]
    \begin{center} 
  \includegraphics[width=8 cm]{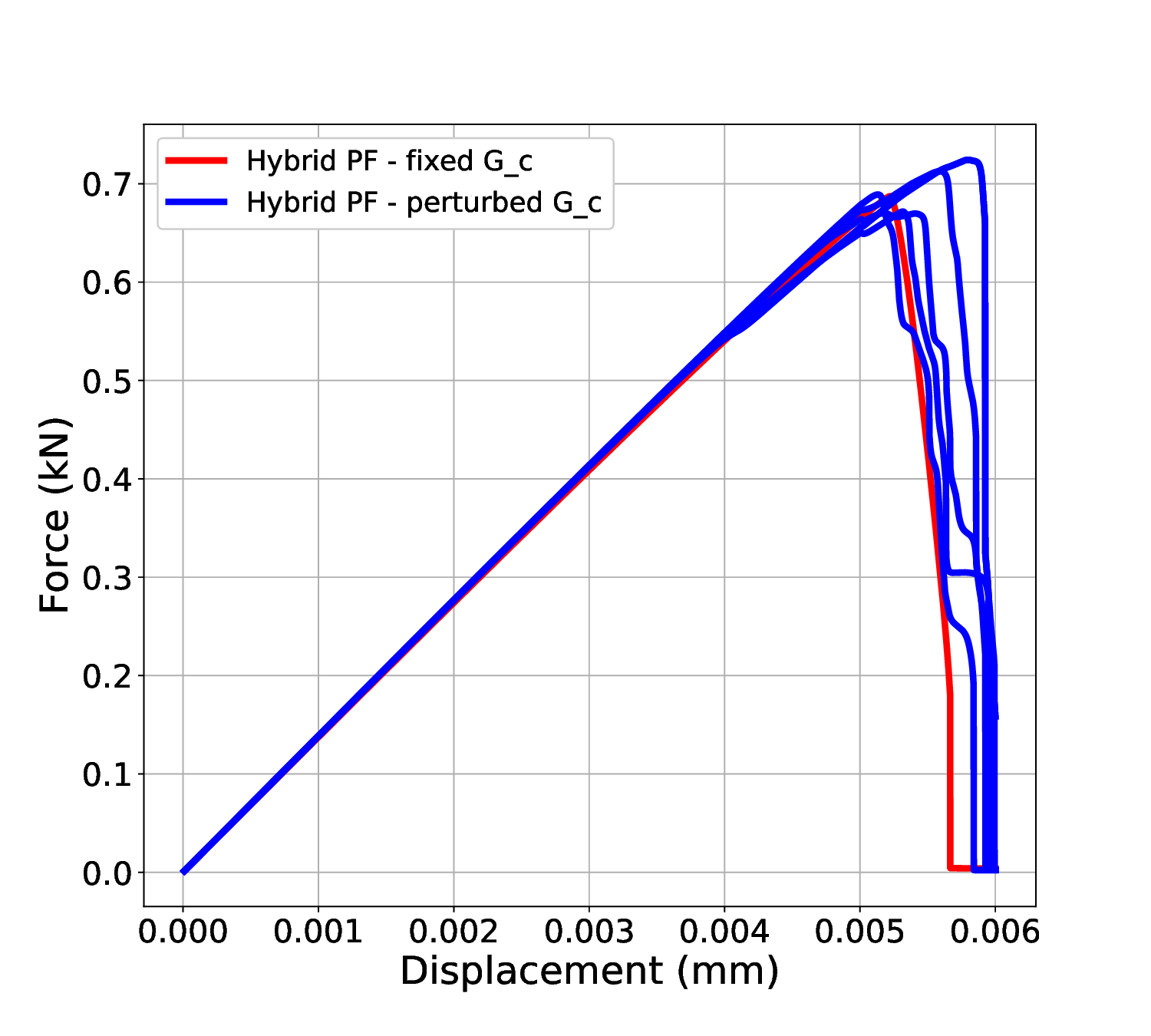}
\end{center} 
\caption{Displacement-force curves of the mode-I fracture of 5 simulations with Mat\'ern covariance (blue lines) compared with the  baseline curve (red line).}\label{sigma_SR_M1} 
\end{figure}

We perform five simulations for each type of fracture using Mat\'ern covariance Gaussian random field realisations. Figure \ref{RM_SR1} shows the propagation path of the initial crack for the pure mode-I fracture. The Figure \ref{frac1} displays the perturbations produced by the random field $G_c$ in contrast to the base trajectory that employs a fixed $G_c$. Figure \ref{sigma_SR_M1} illustrates the comparison between displacement-forces caused by the Gaussian random field employing $G_c$ and the baseline displacement-force curve that employs fixed $G_c$.

\begin{figure*}[h!]
\centering
\begin{subfigure}[b]{.45\linewidth}
\caption{}
\includegraphics[width=7cm]{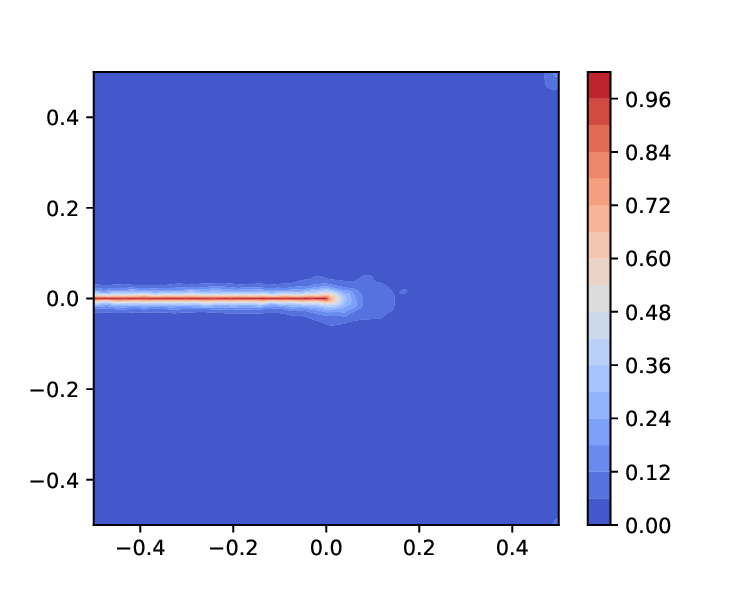}
\end{subfigure}
\begin{subfigure}[b]{0.45\linewidth}
\caption{}
\includegraphics[width=7cm]{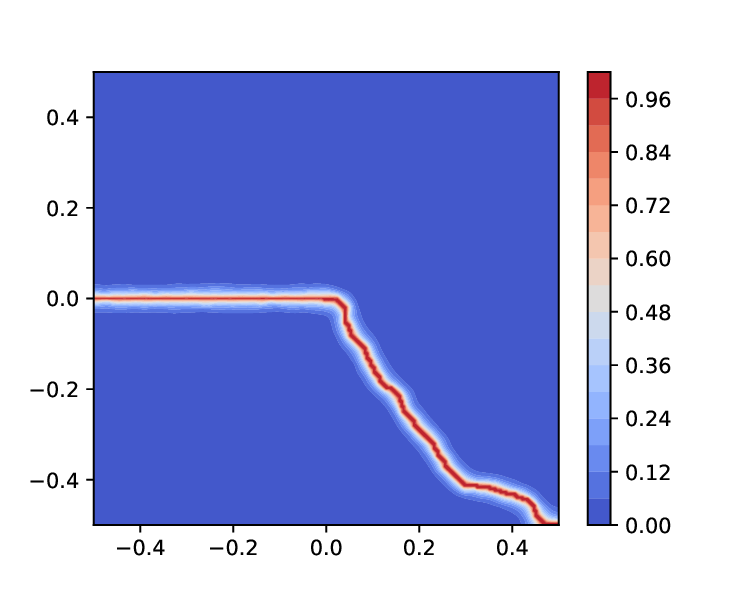}
\end{subfigure}
\caption{Trajectory of a mode-II fracture with the hybrid P-F model at displacement of: a) $5.3\times10^{-3}$ mm; b)  $16\times10^{-3}$ mm.}\label{RM_CR1}
\end{figure*}

\begin{figure}[H]
    \begin{center} 
  \includegraphics[width=8 cm]{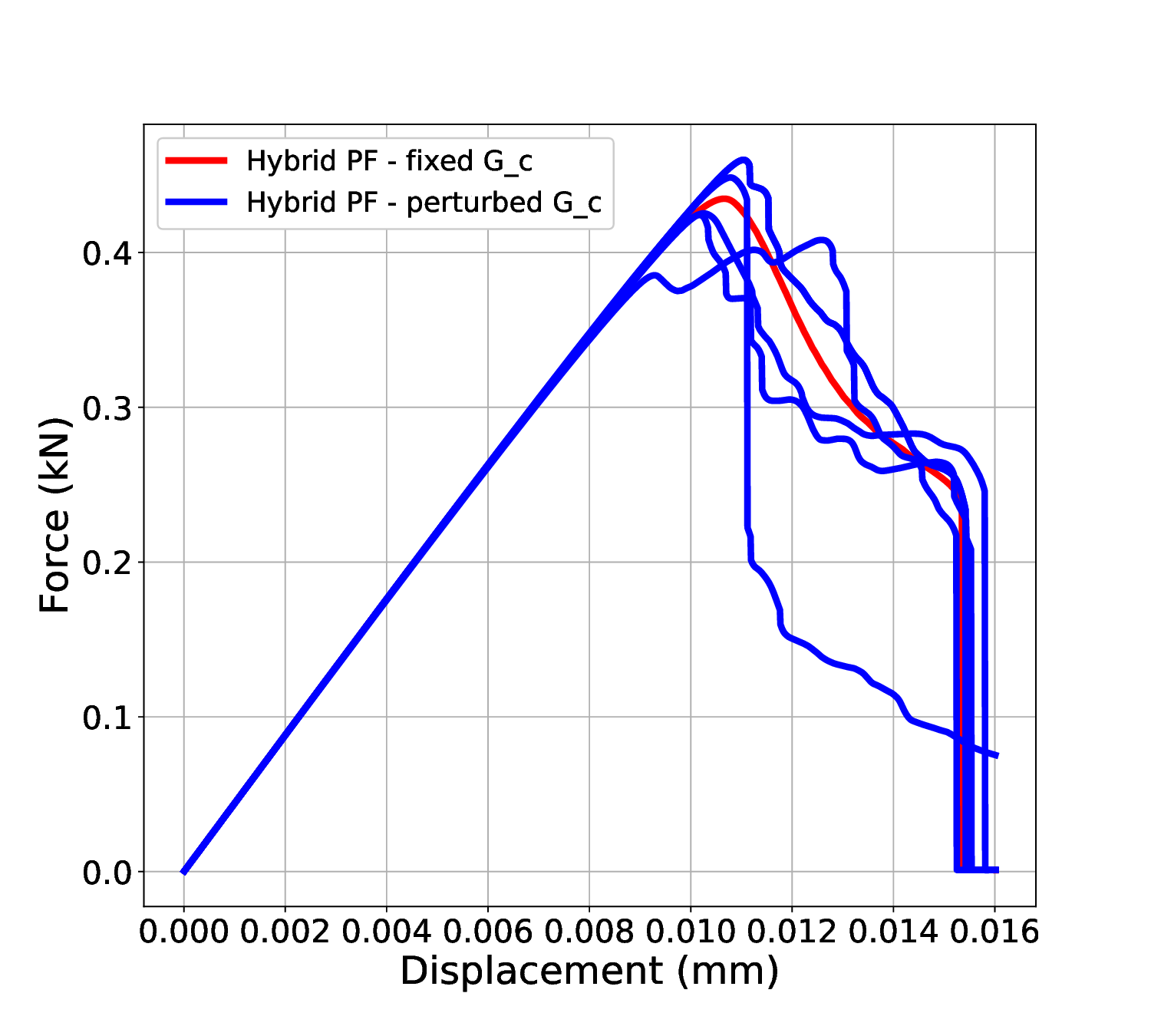}
\end{center} 
\caption{Displacement-force curves of 5 perturbed shear mode-II fractures (blue lines), baseline curve (red line).}\label{sigma_CR_M1}
\end{figure}

\begin{figure*}[h!]
\centering
\begin{subfigure}[b]{.45\linewidth}
\caption{}
\includegraphics[width=7cm]{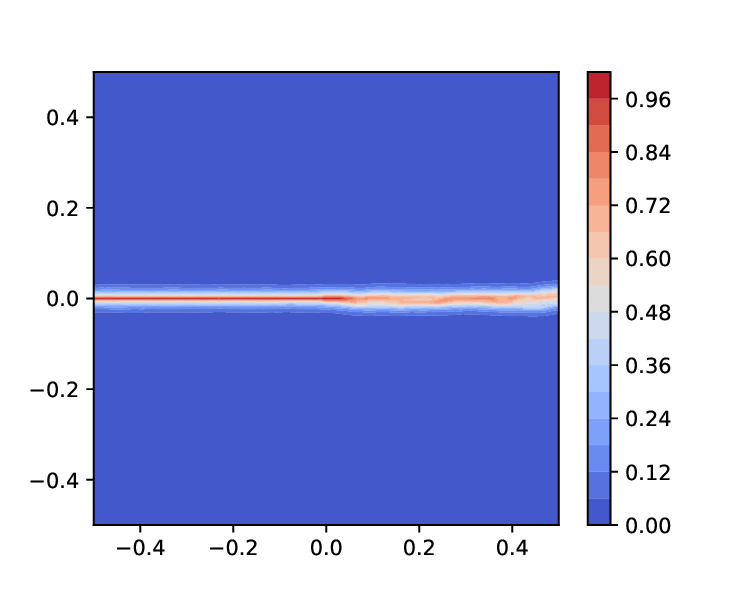}
\end{subfigure}
\begin{subfigure}[b]{0.45\linewidth}
\caption{}
\includegraphics[width=7cm]{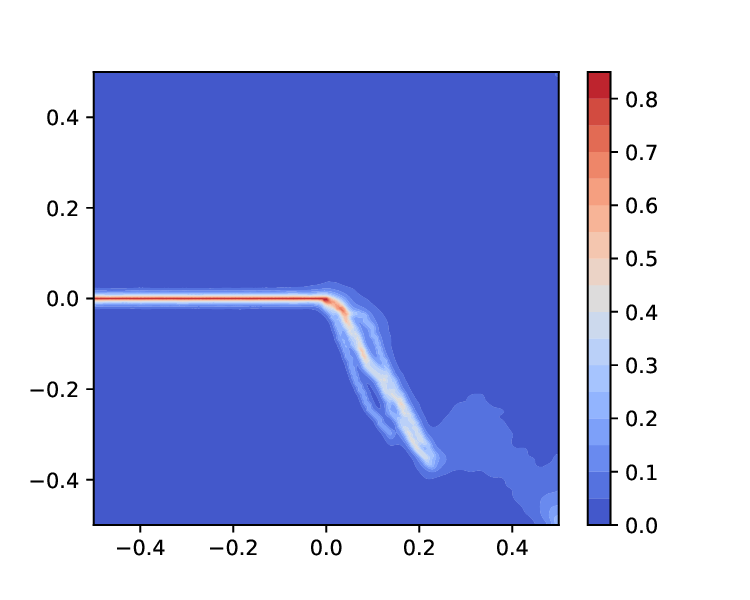}
\end{subfigure}
\caption{Perturbed paths with the hybrid P-F model for: a) mode-I fracture, b) shear mode-II fracture.}\label{over_02}
\end{figure*}

The second set of experiments considers five realisations of a Gaussian random field with Mat\'ern covariance for shear mode II fracture, using displacement increments of $\Delta u$ $=1 \times 10^{-4} \mathrm{mm}$ to $u = 1.6\times 10^{-2}$ mm. In Figure~\ref{RM_CR1} we observe a perturbed path of the crack at the displacement of $u=5.3\times 10^{-3}$ mm, and $u = 1.6\times 10^{-2}$ mm when the crack is propagated, we can compare the result with the path of the propagated crack with fixed $G_c$ in Figure \ref{frac_shear}. The perturbed displacement-force curves associated with the five perturbed cracks from the five samples of the Gaussian random field are shown in Figure ~\ref{sigma_CR_M1}.  We observe different intensities and velocities as the crack is created and propagates.  Figure ~\ref{over_02} shows the overlap of the five possible paths for the cracks in shear mode. 

\begin{figure}[H]
%\begin{multicols}{2}
    \begin{center} 
  \includegraphics[width=7 cm]{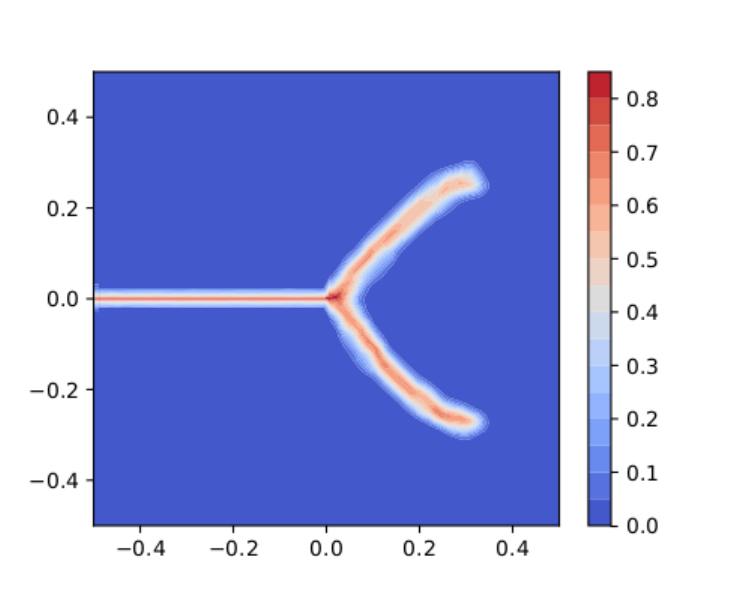}
\end{center} 
\caption{Overlap of 5 perturbed crack paths with the weighted-variational model.}\label{sigma_com_00}
\end{figure}

\begin{figure}[h]
%\begin{multicols}{2}
    \begin{center} 
  \includegraphics[width=8 cm]{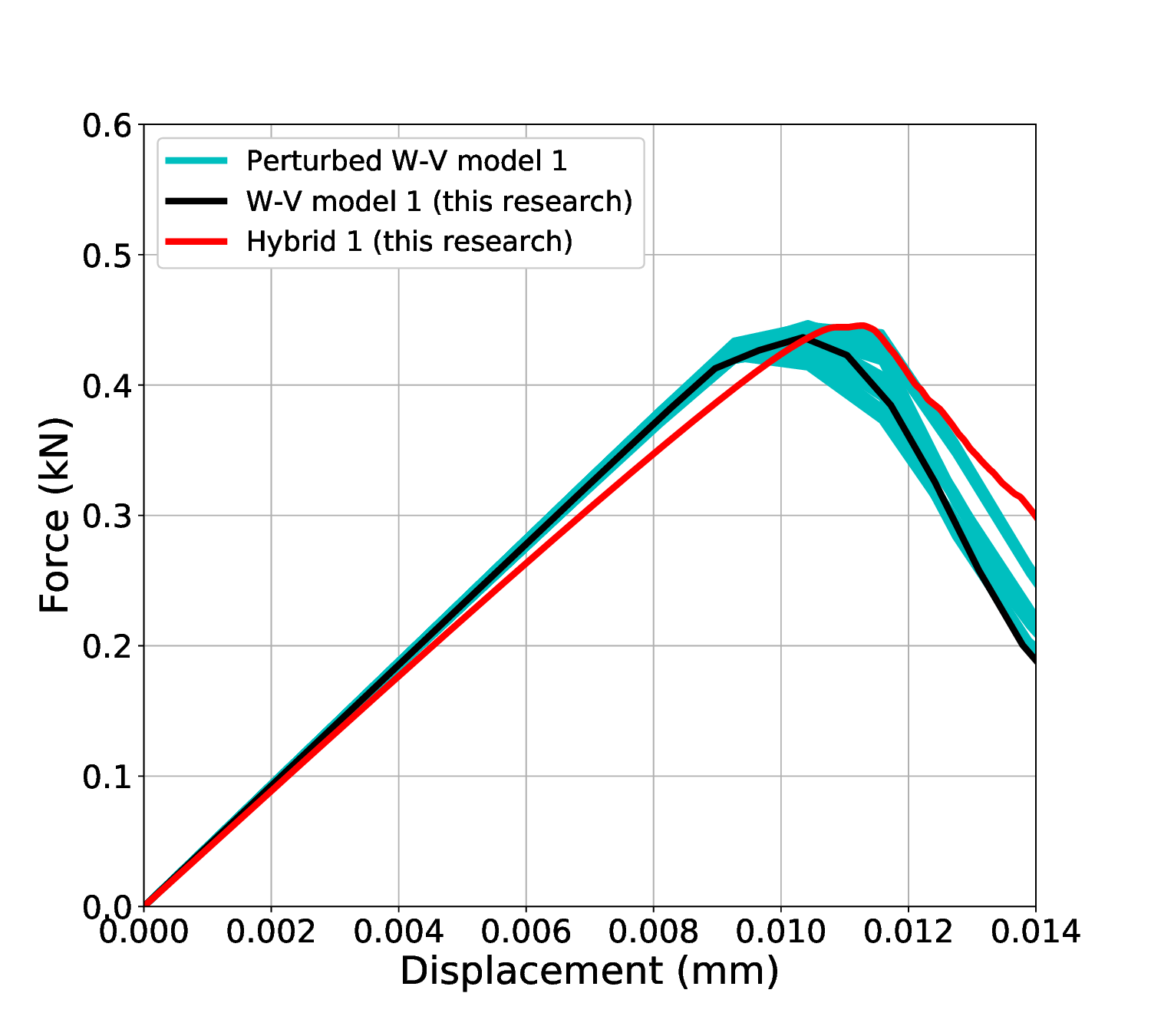}
\end{center} 
\caption{Comparative paths of 20 simulations of the weighted-variational model 1 for the shear mode-II fracture with different Gaussian random field with Mat\'ern covariance.}\label{sigma_com_11}
\end{figure}

The Gaussian random field realizations are employed in weighted-variational model 1 (Perturbed W-V model 1). Figure \ref{sigma_com_00} depicts the overlap of 5 crack-propagation paths that correspond to five samples for $G_c$. Figure \ref{sigma_com_11} shows the displacement-force curve.  

Simulation results have shown that even after including Gaussian perturbations in the parameter $G_c$, our model reproduces crack propagation behaviour similar to the hybrid phase field model. Figure \ref{sigma_com_11} illustrates the overlap of the displacement-force curves from 20 samples of the perturbed weighted-variational model 1 with the simulations from our Hybrid Phase field 1 (Hybrid 1) and W-V model 1.

\begin{figure}[H]
%\begin{multicols}{2}
    \begin{center} 
  \includegraphics[width=8 cm]{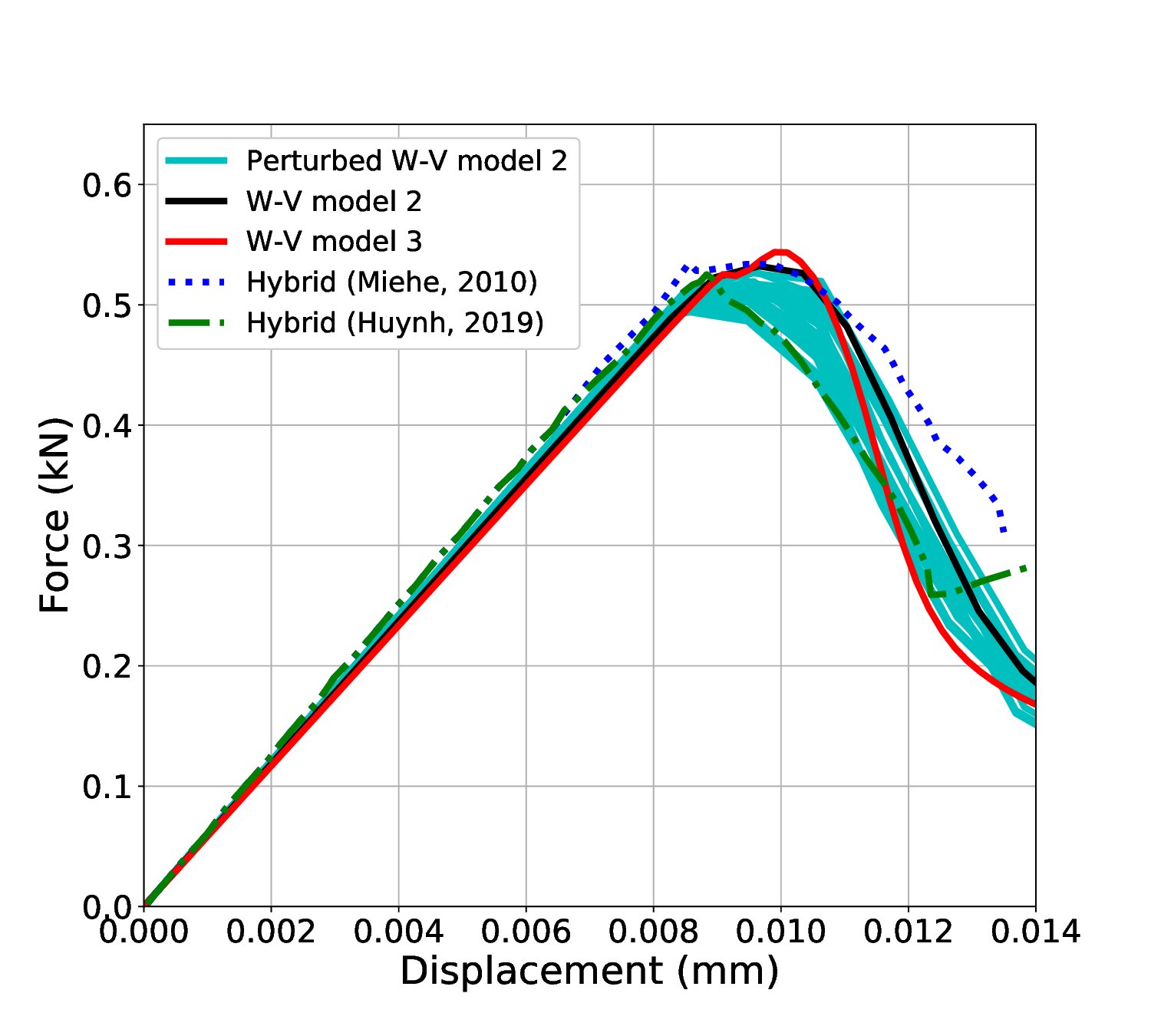}
\end{center} 
\caption{Comparative paths of 20 simulations of the weighted-variational model 2 for the shear mode-II fracture with different Gaussian random field with Mat\'ern covariance.}\label{sigma_com_111}
\end{figure}

Figure \ref{sigma_com_111} shows 20 perturbed samples of the weighted-variational model 2 and compares the outcomes with the profiles presented in \cite{MIEHE_2010a} and \cite{HUYNH_2019}, as well with our results of W-V model 2 and W-V model 3.
The hybrid phase-field model has a mean execution time of approximately 7 hours, while the variational model has a mean execution time of around 20 minutes. This represents a time complexity reduction of approximately 90\%, resulting in lower computational time, reduced storage and memory operations.
Only a few simulations were executed in these calibration tests.  The next section presents a comprehensive Monte Carlo study and compares it with a laboratory experiment. 

\subsection{Comparisons with a laboratory experiment and Monte Carlo study}

\begin{center}
\begin{figure*}[h!]
\centering
\begin{subfigure}[b]{.3\linewidth}
\caption{}
\includegraphics[width=4cm]{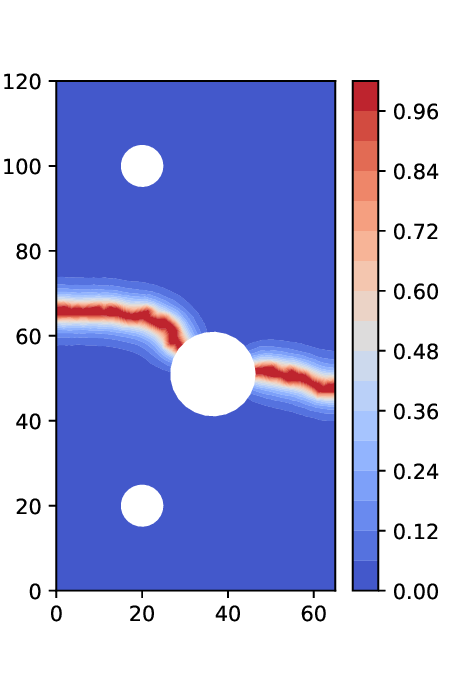}
\end{subfigure}
\begin{subfigure}[b]{0.15\linewidth}
\caption{}
\includegraphics[width=4cm]{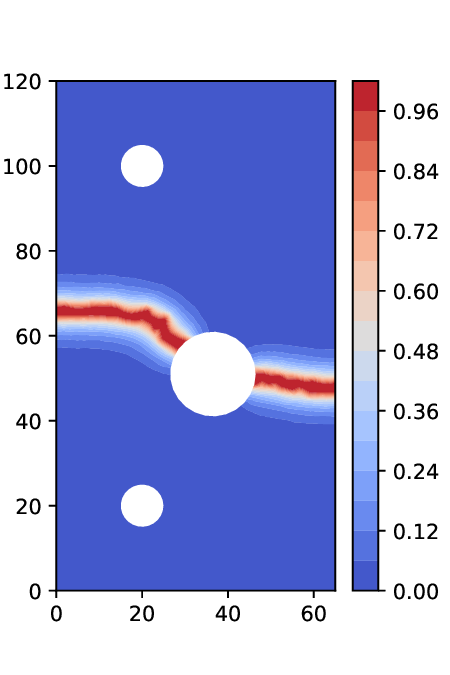}
\end{subfigure}
\caption{Comparative results of displacement-force: a) phase field model - 80 points ($\Delta u = 0.0275$), $\approx$ 45 minutes, b) weighted-variational model - 50 points ($\Delta u = 0.044$), $\approx$ 5 minutes.}\label{sigma_3exp}
\end{figure*} 
\end{center}

\begin{figure}[H]
    \begin{center} 
  \includegraphics[width=9 cm]{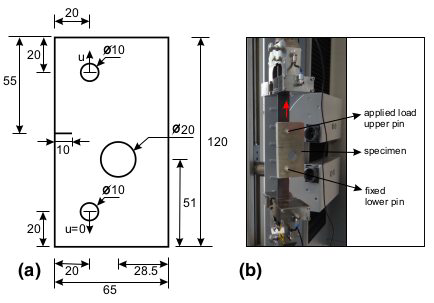}
\end{center} 
\caption{Brittle fracture experiment setting, (a) cement specimen design (mm) and (b) loads applied; figure taken from \cite{AMBATI_2015}. }\label{sigma_exp_2}
\end{figure}

\begin{figure}[H]
    \begin{center} 
  \includegraphics[width=7 cm]{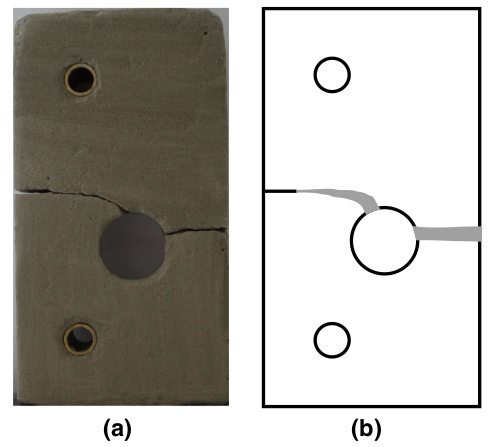}
\end{center} 
\caption{Brittle fracture experiment, (a) cement specimen after fracture and (b) expected fracture path; figure taken from \cite{AMBATI_2015}. }\label{sigma_exp_3}
\end{figure}

To test how realistic our approach is in simulating real brittle fractures, we apply the hybrid phase field and weighted variational surrogate models, with $G_c$ as a Gaussian random field, and try to mimic the experimental results presented in Ambati {\sl et al.}~\cite{AMBATI_2015} (see Figures~\ref{sigma_exp_2} and~\ref{sigma_exp_3}). The geometry and boundary conditions are shown in figure~\ref{sigma_exp_2}(a). The specimen was made of cement mortar composed of $22$\% cement (Cement I 32.5: high alumina cement 4:1), $66$\% sand (grain size $< 1$~mm), and $12$\% water, resulting in a water ratio of $0.55$ (for more details see~\cite{AMBATI_2015}).  The specimen is a notched plate loaded by an upper pin and a fixed lower pin.  The specimen has a hole offset from the center to induce mixed mode fracture.  The displacement controlled loading was $0.1$~mm/min, see figure \ref{sigma_exp_2}(b).  

The material parameters, taken from \cite{AMBATI_2015}, are $\lambda = 1.94$~kN/mm$^2$, $\mu = 2.45$~kN/mm$^2$, $G_c = 2.8 \times 10^{-3}$~kN/mm. Figure \ref{sigma_exp_3} shows the physical experimental with the resulting crack and the expected fracture presented in \cite{AMBATI_2015}.

The simulations performed in \cite{AMBATI_2015} used $\ell = 0.1$ mm, with 25085 quadrilateral elements for the finite element method in the hybrid phase field model. To reduce the computational cost and following the comments in section \ref{sec:4}, in our simulations we discretized the domain $\Omega$ into $4,947$ triangular elements and used linear test functions for the FEM formulation, and we set $\ell = 2 h = 2.8008$, which is $h= 1.1004$. A simulation of fracture propagation using the hybrid phase field model with 80 displacement increments ($\Delta u = 0.0275$) takes about 45 minutes to complete, while the weighted variational surrogate model with $\xi=1.75$ and 50 displacement increments ($\Delta u = 0.044$) has a run time of about 5 minutes.  The surrogate model is comparable and describes the expected behavior of the crack, as shown in Figure~\ref{sigma_3exp}.

These results clearly show that the weighted-variational model, calibrated with the appropriate set of parameters ($\ell$, $\Delta_u$ and mesh size), can accurately reproduce the results of the hybrid phase-field model. In our simulations, in both models, the first crack is generated around the applied load of $0.39$~mm, similar to the actual numerical experiment where the first crack appears at a load of $\approx. 0.4$ mm, see \cite{HUYNH_2019}. 

\begin{center}
\begin{figure*}[h!]
\centering
\begin{subfigure}[b]{.3\linewidth}
\caption{}
\includegraphics[width=4cm]{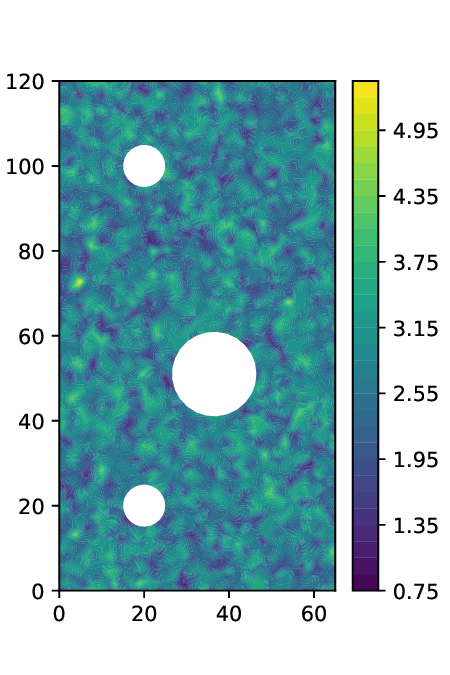}
\end{subfigure}
\begin{subfigure}[b]{0.15\linewidth}
\caption{}
\includegraphics[width=4cm]{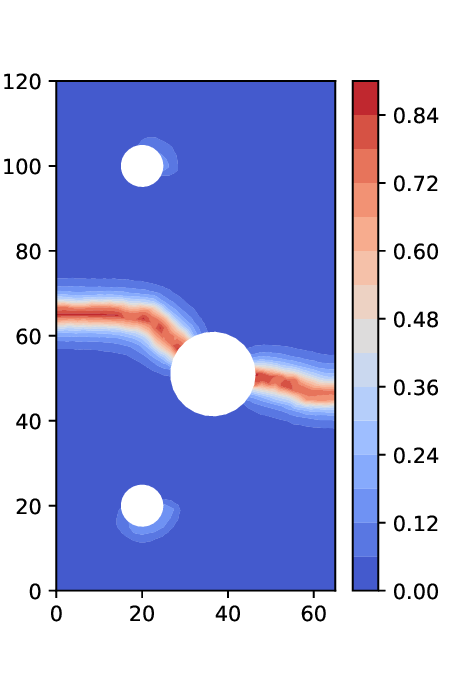}
\end{subfigure}
\caption{a) Gaussian random field with Mat\'ern covariance; b)
  Overlap of the 500 simulations of the perturbed fractures.}\label{GRF_con}
\end{figure*} 
\end{center}

\begin{center}
\begin{figure*}[h!]
\centering
\begin{subfigure}[b]{.5\linewidth}
\caption{}
\includegraphics[width=7cm]{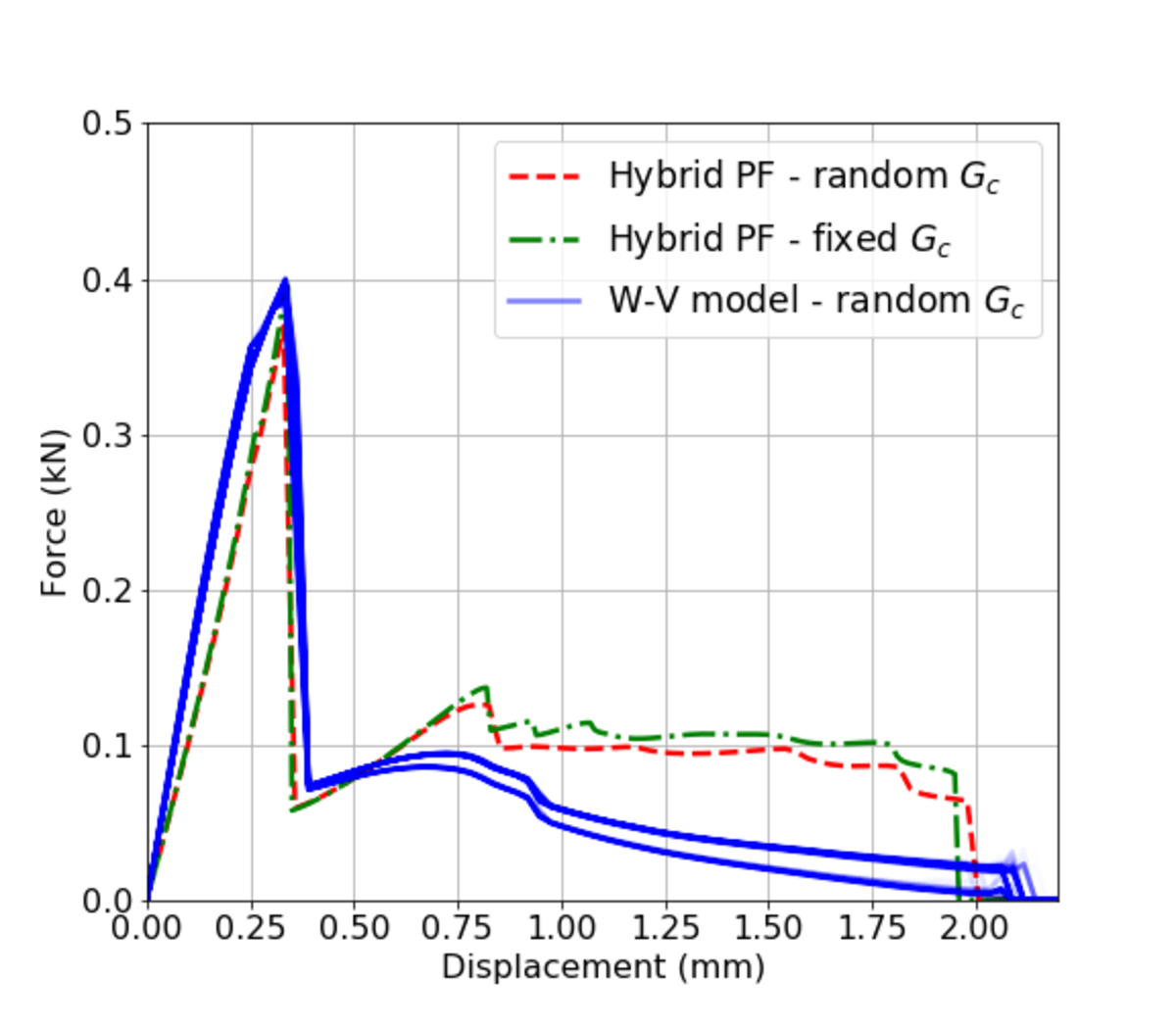}
\end{subfigure}
\begin{subfigure}[b]{0.3\linewidth}
\caption{}
\includegraphics[width=7cm]{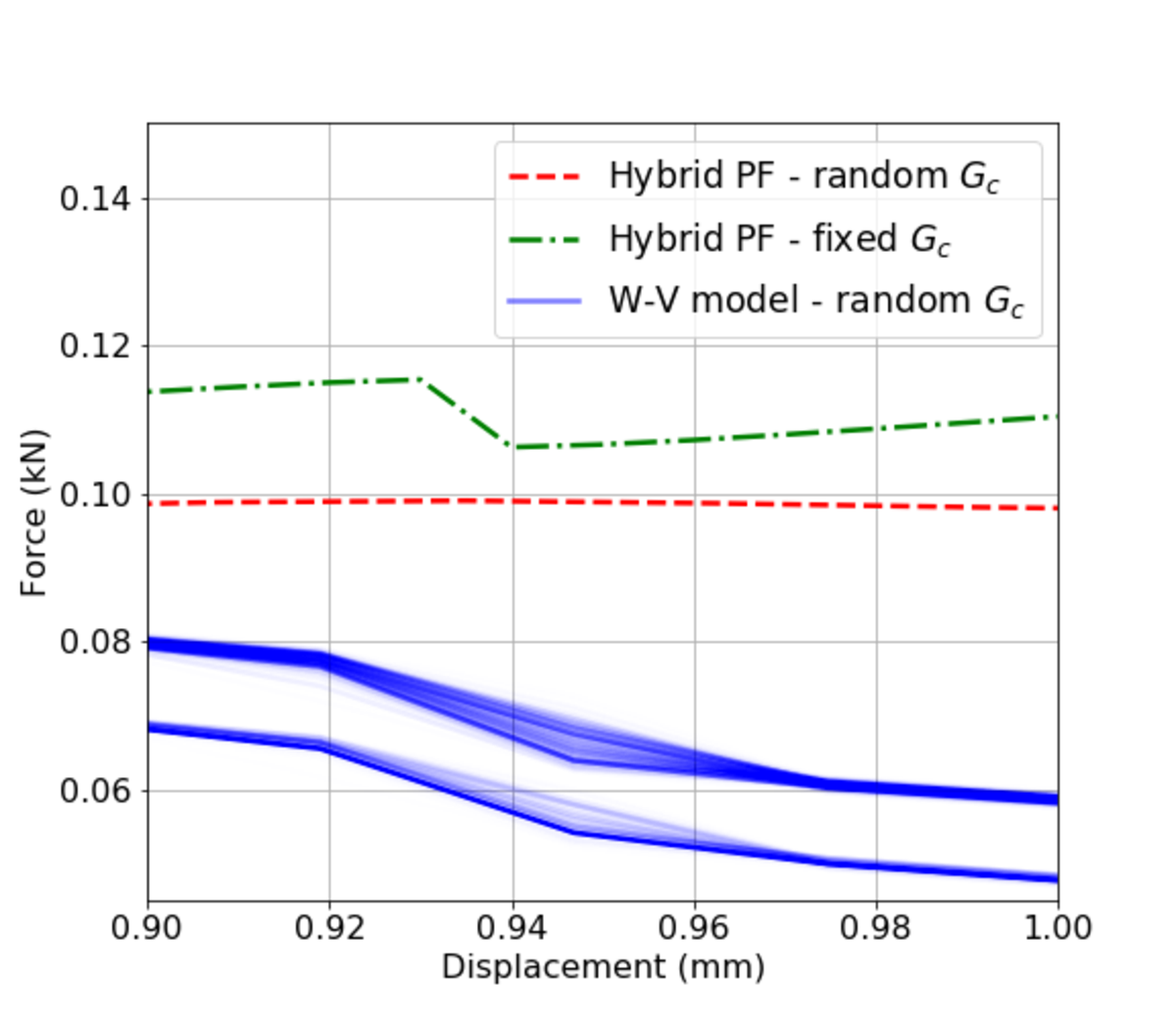}
\end{subfigure}
\caption{ a) Comparative displacement-force curves obtained with the Hybrid PF with fixed $G_c$ (point-dashed green line), the Hybrid PF with random $G_c$ (dashed red line), and the overlap of curves from 500 simulations of the W-V model with random $G_c$. b) Zoom in $[0.9, 1.0]\times[0.045, 0.090]$.}\label{GRF_cuantiles}
\end{figure*} 
\end{center}

\begin{center}
\begin{figure*}[h!]
\centering
\begin{subfigure}[b]{.5\linewidth}
\caption{}
\includegraphics[width=7cm]{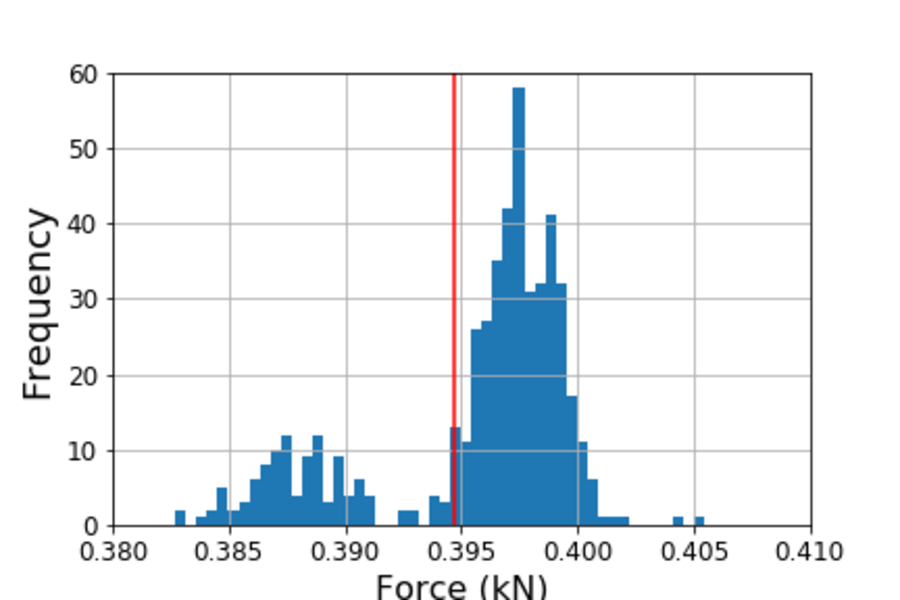}
\end{subfigure}
\begin{subfigure}[b]{0.3\linewidth}
\caption{}
\includegraphics[width=7cm]{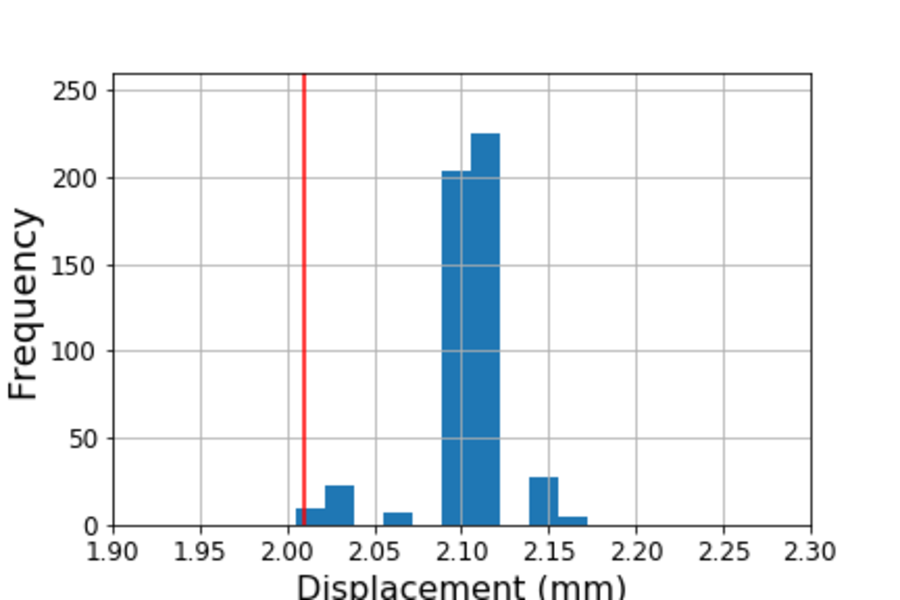}
\end{subfigure}
\caption{a) Histogram of the maximum values of the force (kN) of the displacement-force curves of the 500 simulations with the W-V model with random $G_c$ compared to the maximum value of the force of the hybrid PF with fixed $G_c$ (red line). b) Histogram of the displacement where the final crack is obtained with the W-V model with random $G_c$ compared to a red line corresponding to the generated final crack of the hybrid PF - fixed $G_c$.}\label{GRG_hist}
\end{figure*} 
\end{center}

\subsubsection{Results of simulations with small perturbations of $G_c$}\label{sec:R1}

To simulate inhomogeneities in the material and to obtain more realistic simulations, we performed a Monte Carlo study with 500 simulations corresponding to 500 different realizations of $G_c$, modeled as a Gaussian random field, as explained in the previous section \ref{sec:3}. The Gaussian field for $G_c$ has parameters to simulate a mean $G_c$ perturbed with random inhomogeneities, the range of perturbation of $G_c$ is 10\%. Figure~\ref{GRF_con} shows the overlap of the resulting 500 perturbed fracture trajectories. The results are quite similar to the actual experimental results, as seen in Figure~\ref{sigma_exp_3}.

 Comparative displacement-force curves are shown in Figure \ref{GRF_cuantiles}, the curves are obtained from simulations with the hybrid phase-field method with a fixed value of $G_c$ (hybrid PF - fixed $G_c$, dotted green line), with the hybrid phase-field method with random $G_c$ (hybrid PF - random $G_c$, dashed red line), and with the 500 simulations with the weighted variational model with random $G_c$ (W-V model). In this plot, two different profiles of the displacement-force curves resulting from the random perturbations in $G_c$ can be seen.

We also present a plotted zoomed area of the displacement-force curves in the window of $[0.9, 1.0]\times[0.045, 0.15]$ in figure \ref{GRF_cuantiles}-b).  In this figure we can see the different profiles or behaviors generated by the perturbed $G_c$. In figure \ref{GRF_cuantiles}-a), we observe that the profiles show that the first crack appears around the displacement of 0.39 mm. On the other hand, a histogram showing the distribution of the maximum peaks in the displacement force curves of the 500 simulations of W-V is presented in figure \ref{GRG_hist}-a).  The red line represents the value of the maximum peak of the displacement-force curve obtained by the hybrid phase-field method with a fixed value of $G_c$ (hybrid PF - fixed $G_c$), this value corresponds to $\approx 0.395$ kN. In addition, the final crack is obtained around 2 mm in all simulations, as shown in the histogram of final cracks, Figure \ref{GRG_hist}-b).

\subsubsection{Results of simulations with high perturbations of $G_c$}

To analyze our proposal, we consider an extreme case where we generate a high perturbation for $G_c$, and we used the maximum perturbation of $\approx 100\%$, these simulations allow us to obtain more information about the cracks in extreme conditions. We ran a new set of 500 simulations, corresponding to 500 realizations of Gaussian random fields with this new scenario.

\begin{center}
\begin{figure*}[h!]
\centering
\begin{subfigure}[b]{.5\linewidth}
\caption{}
\includegraphics[width=7cm]{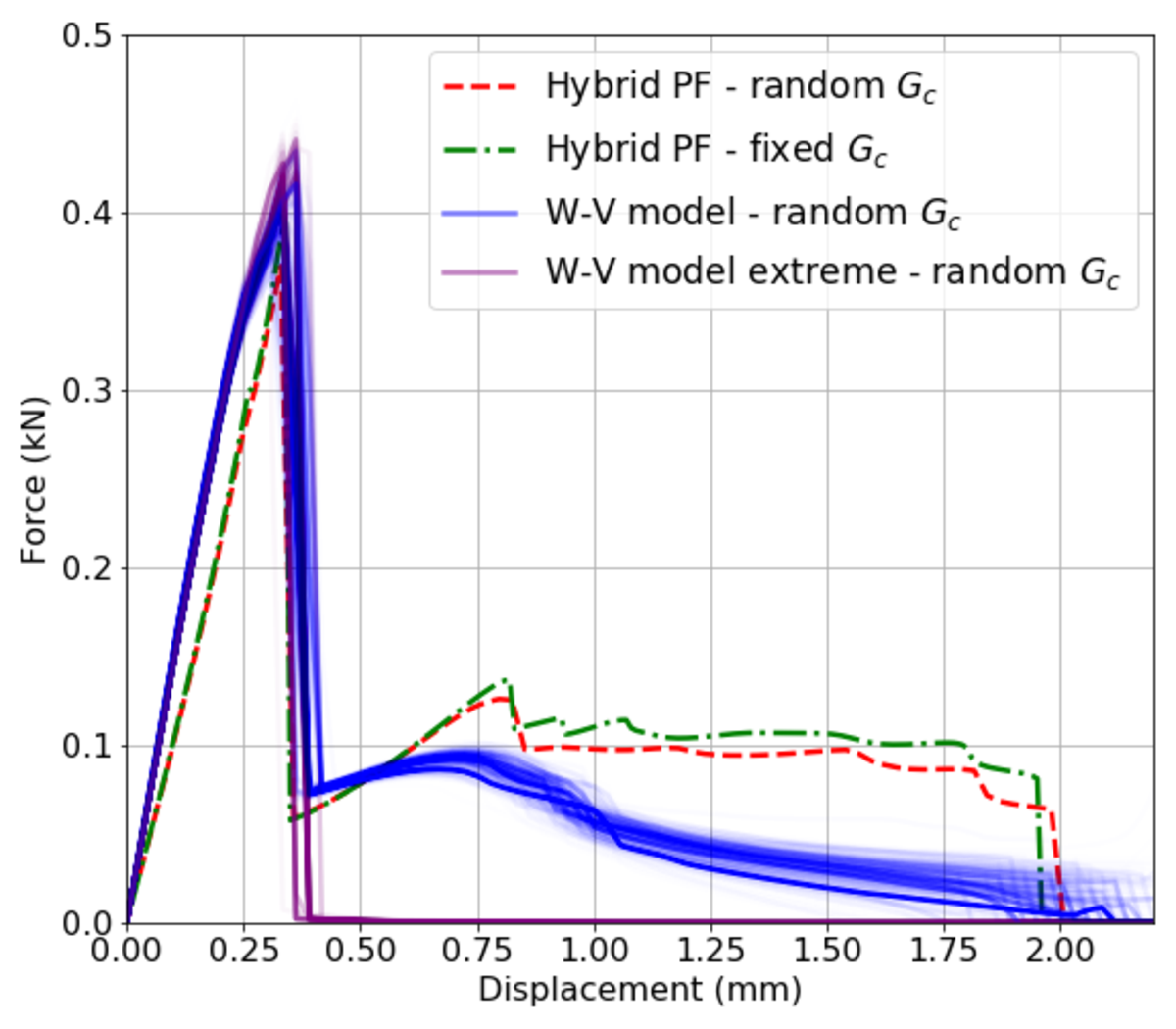}
\end{subfigure}
\begin{subfigure}[b]{0.3\linewidth}
\caption{}
\includegraphics[width=7cm]{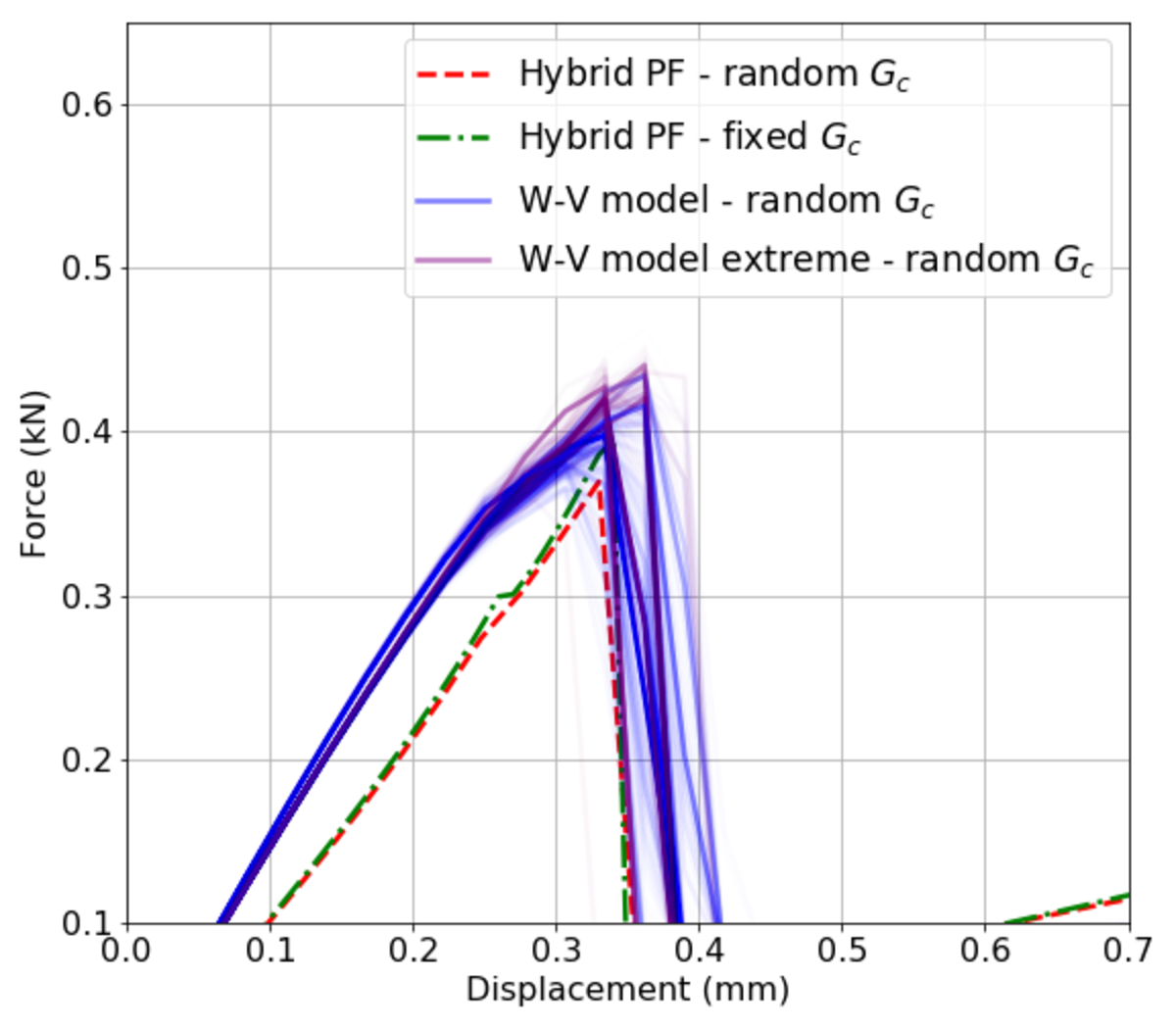}
\end{subfigure}
\caption{a) Comparative displacement force curves obtained with the Hybrid PF with fixed $G_c$ (dotted green line), the Hybrid PF with random $G_c$ (dashed red line), and the overlap of curves from 500 simulations of the W-V model with random $G_c$ (blue lines correspond to expected profiles, and purple lines correspond to extreme fractures). b) Zoom in $[0.0, 0.7]\times[0.1, 0.6]$. }\label{GRF_cuantiles3}
\end{figure*} 
\end{center}

\begin{center}
\begin{figure*}[h!]
\centering
\begin{subfigure}[b]{.3\linewidth}
\caption{}
\includegraphics[width=4cm]{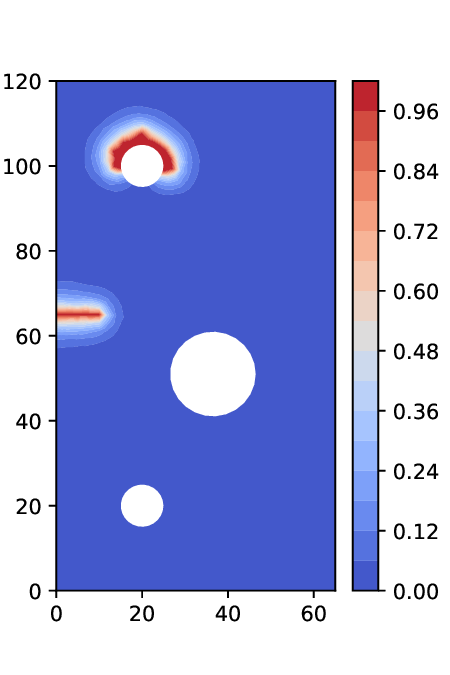}
\end{subfigure}
\begin{subfigure}[b]{0.15\linewidth}
\caption{}
\includegraphics[width=4cm]{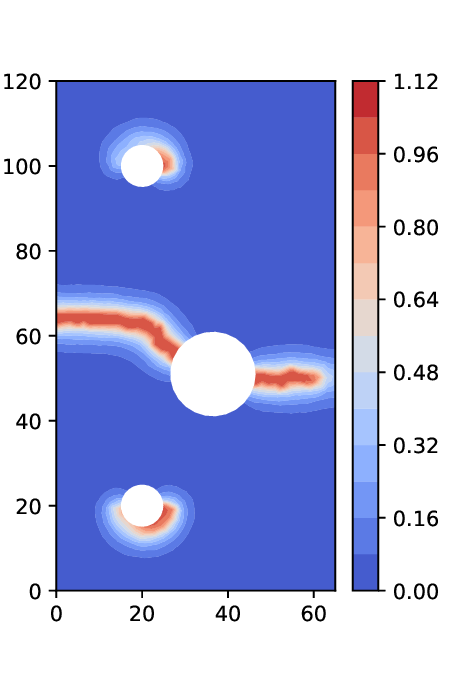}
\end{subfigure}
\caption{Results of cracks with high perturbation in $G_c$: a) Extreme simulation of the W-V model; b) Extreme simulation of the phase-field method.}\label{GRF_cuantiles2}
\end{figure*} 
\end{center}

In Figure \ref{GRF_cuantiles3}, we show the overlap of the displacement force curves obtained from simulations of the Hybrid Phase Field method with a fixed value of $G_c$ (Hybrid PF - fixed $G_c$, dotted green line), with the hybrid phase-field method with random $G_c$ (hybrid PF - random $G_c$, dashed red line), and with the 500 simulations with the weighted variational model with a high Gaussian perturbation of $G_c$ (W-V model).  For this simulation we obtained two main scenarios, the first scenario produces the already seen profile for the crack with small perturbations, as in the case of subsection \ref{sec:R1}.  The second, unexpected, scenario is a strange \textit{incomplete} fracture that produces a crack in the upper hole of the specimen, see figure \ref{GRF_cuantiles2}-a).  These scenarios can be observed in the profiles of the displacement force curves in Figure \ref{GRF_cuantiles3}, of the 500 simulations, 433 (86\%) correspond to the first scenario (blue lines) and 66 (13.4\%) correspond to the second scenario (purple lines). We also plotted the histograms of the maximum force value of the simulations and the final displacement in Figure \ref{GRF_cuantiles2}.

\begin{center}
\begin{figure*}[h!]
\centering
\begin{subfigure}[b]{.5\linewidth}
\caption{}
\includegraphics[width=7cm]{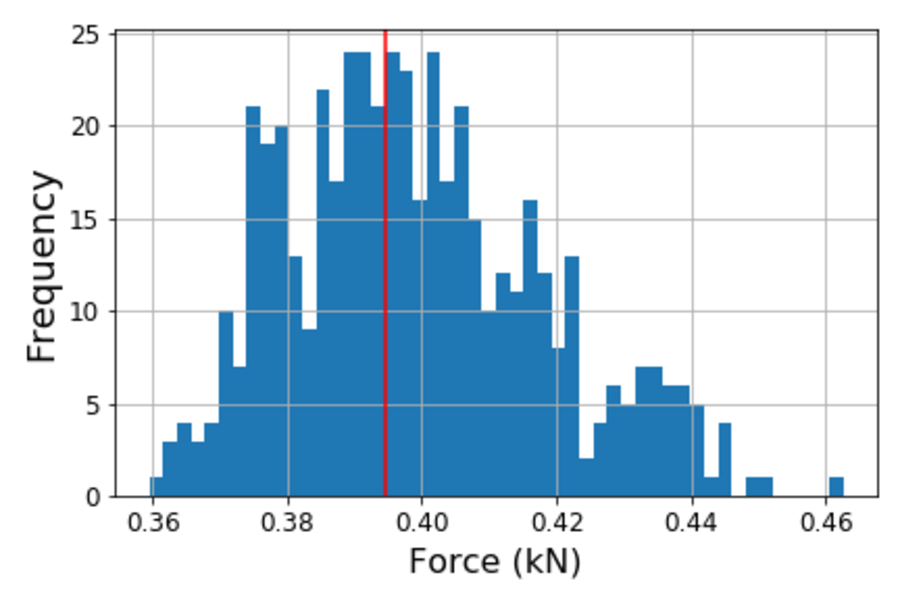}
\end{subfigure}
\begin{subfigure}[b]{0.3\linewidth}
\caption{}
\includegraphics[width=7cm]{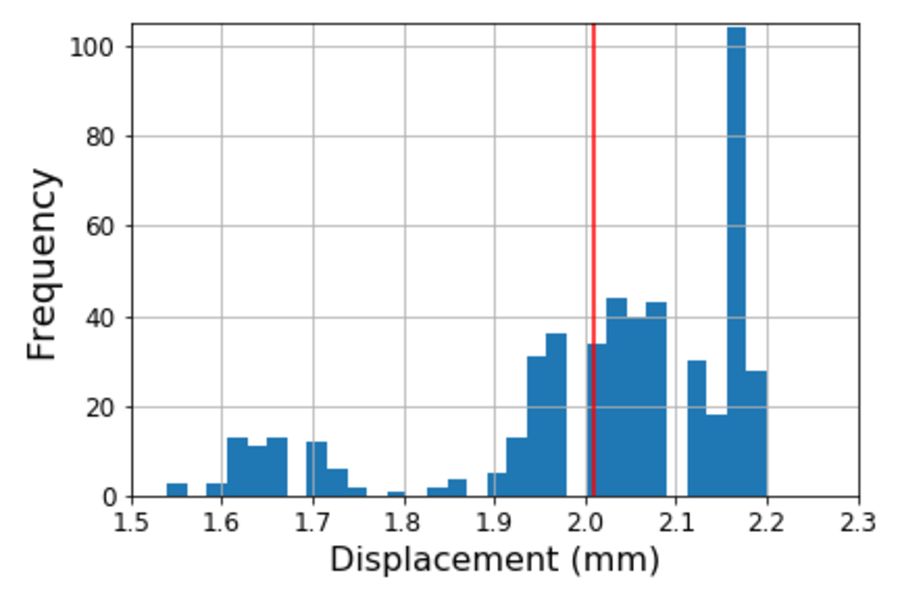}
\end{subfigure}
\caption{a) Histogram of the maximum values of the force (kN) of the displacement-force curves of the 500 simulations with the W-V model with high random $G_c$ compared to the maximum value of the force of the hybrid PF with fixed $G_c$ (red line). b) Histogram of the displacement where the final crack is obtained with the W-V model with random $G_c$ compared to a red line corresponding to the generated final crack of the hybrid PF - fixed $G_c$.}\label{GRF_cuantiles4}
\end{figure*} 
\end{center}

To further investigate the strange crack obtained in 13\% of the simulations of the Monte Carlo study (using the W-V model), we performed a simulation of the phase field model with the same mesh and a realization of $G_c$ that produces this unexpected crack.  Using the computationally expensive phase field model, we observe a similar but not identical scenario where the crack is created in the holes of the specimen and the trajectory of crack propagation is not complete, Figure \ref{GRF_cuantiles2}-b). In fact, the appearance of this new crack is due to the high perturbation of $G_c$, and about 13\% of the cases will follow this incomplete crack, breaking the specimen in the joint holes.  An evaluation of the phase field model seems to confirm this possibility, i.e. it is not a mere fiction of our surrogate W-V, although both simulations are not identical.  However, this is a result of our Monte Carlo study, which introduces heterogeneity in $G_c$, and this alternative cracking scenario would have remained unknown without the availability of a fast surrogate, namely our W-V model.  The entire 500 simulations using the W-V model took $\approx 42$ hours to compute (using 24 processors running at 2.10 GHz), whereas if we had used the phase field model, these 500 simulations would have taken $\approx 16$ days to compute using the same computing power; including if we increase the number of elements as standard simulations in the literature, the Monte Carlo study would take several months.

\begin{figure}[H]
    \begin{center} 
  \includegraphics[width=8 cm]{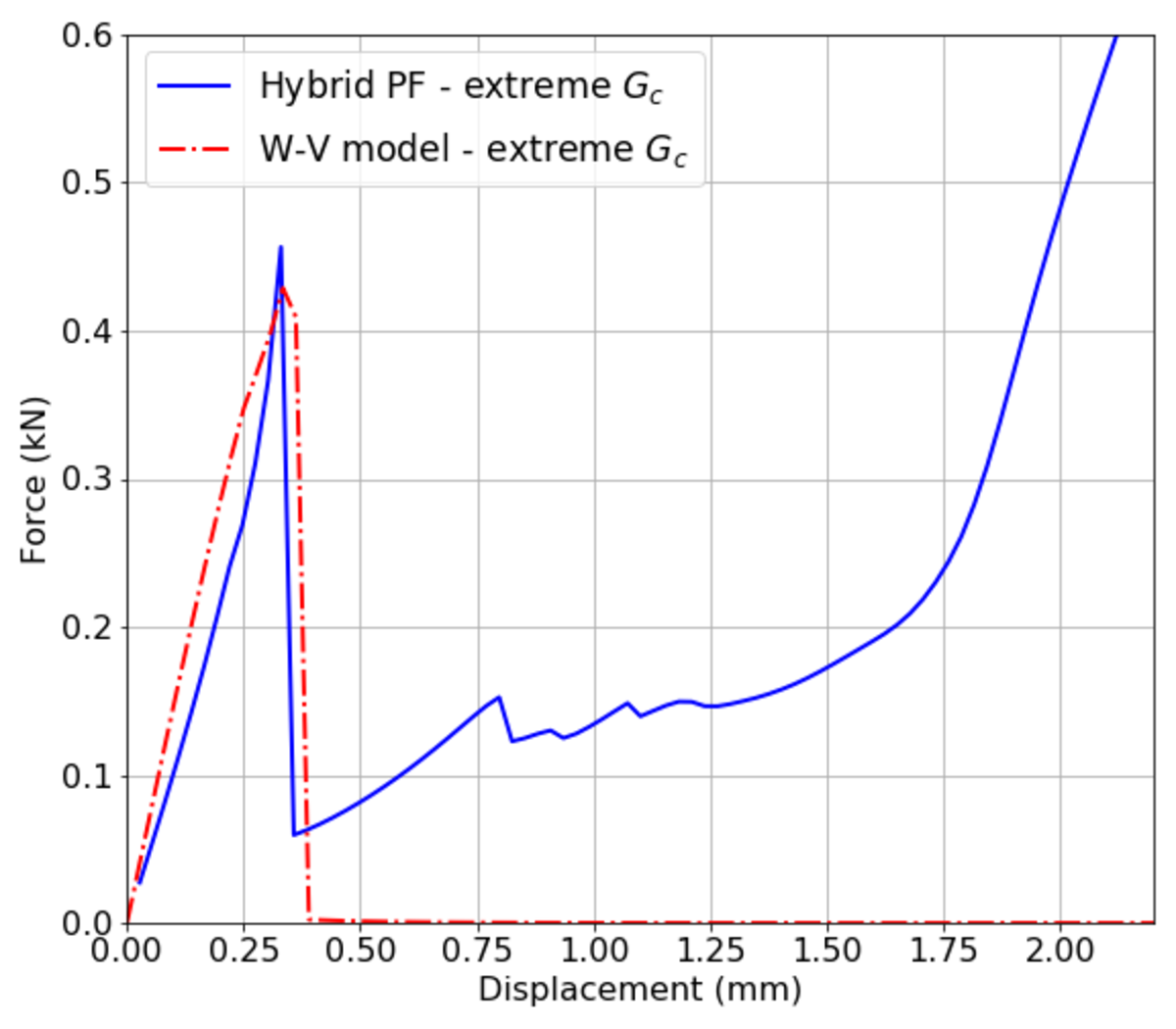}
\end{center} 
\caption{Comparison of perturbed displacement-force curves 
  with the hybrid phase field method (blue solid line) and the weighted variational model (red dashed line).}\label{GRG_hist2}
\end{figure}

The displacement force curve of a hybrid phase field method and of the W-V model (surrogate model) are shown in \ref{GRG_hist2}, in both cases we obtain extreme profiles corresponding to the two different scenarios of crack propagation just mentioned.
%This behavior was possible from the previous information by the surrogate proposed model.

%In real structures (concrete, ceramic, etc.) the physical properties are essential to obtain resistivity and durability.  For this reason, the study of the random properties in materials that can induce a fracture propagation is important for practitioners.The weighted-variational model as a surrogate model of the hybrid phase-field presents an alternative to obtaining simulations in a short time, reducing complexity, and computational storage. We have observed that we recover the expected trajectories of the fractures in our numerical examples.More comprehensive sample sizes, for a Monte Carlo study, are now feasible, using our proposed weighted-variational surrogate model, with execution times of $\approx 5$ min for this experimental example, due to numerical experiments using random fields providing more realistic information about the cracks. 

\section{Conclusion}\label{sec:6}

In this work, we propose the use of the weighted variational model as a surrogate for the hybrid phase-field model to reduce the computational time and memory requirements for simulations.  We have observed the parameter dependence of $\xi$, $\ell$, $\Delta_u$ and mesh size $h$ on the simulations. The calibration of these numerical parameters is necessary to obtain realistic results comparable to real experiments. The two models are compared and a drastic reduction in execution time is achieved, from 45 min to $\approx 5$ min, almost an order of magnitude reduction. 
This opens up the possibility of performing a Monte Carlo study considering a Gaussian random field to model a non-homogeneous critical energy release rate $G_c$.

The use of random fields to analyze the uncertainty in fracture is a topic that should be analyzed to create new materials and study their strength and resistance. Each possible crack trajectory is affected by the random nature of the specimen caused by water, air bubbles, fibers, etc. These anomalies can be described by random fields to simulate more realistic properties of materials. Numerical experiments can provide information to describe crack propagation. This work provides an overview of the use of random fields to analyze the heterogeneous properties of a material. 

In future work, we will study additional physical properties that may
also be statistically descriptive of brittle fractures, and we will analyze the uncertainty quantification of the elastic parameters using a Gaussian random field with Mat\'ern covariance in a Monte Carlo simulation, since these parameters are also affected by the random nature of the specimen. As mentioned above, the main idea of having a computationally cheap surrogate is to allow a Monte Carlo (MC) study of a Gaussian random field model for the critical energy release rate $G_c$ (see section~\ref{sec:3}). That is, special simulations of the stochastic $G_c$ are used to run the model and simulate various crack propagations and their characteristics (mainly the displacement/force curves).  This is only remotely feasible with a fast model that can run within a few minutes to allow at least several hundred MC evaluations \cite{molina2022}, similar to previous work in \cite{YANG_2009,SU_2010,Dsouza_2021}.      

Although equivalent to the above, but a mathematically far more elegant approach is to consider~(\ref{sis_1}) or~(\ref{eq21}) itself as stochastic, basically now aiming to find numerically the multidimensional probability law governing the now stochastic crack propagation etc. see for example \cite{babuska_2005,babuska_2010}.  Although mathematically elegant, this approach adds several orders of magnitude of numerical complexity to an already computationally demanding problem.   
For this reason, MC approaches are preferred in this and a large number of probabilistic studies. 

\section*{Acknowledgments}
 This research is partially founded by CONACYT CB-2016-01-284451 and RDECOM grants, and by ONRG-AFOSR N62909-24-1-2016 grant.

 \section{Data Availability Statement}

Complementary data is available at:\url{https://github.com/pemfc-team-CMCGCY/surrogate_brittle_fractures.git} 

\section*{References}
%\bibliography{reference1}

\end{document}